\documentclass[12pt]{article}
\usepackage{jheppub}
\newcommand{\co}{{\cal O}}

\def\bal#1\eal{\begin{align}#1\end{align}}
\def\alp[#1]{\begin{align}#1\end{align}}

\def\secnum[#1]{\texorpdfstring{$#1$}{TEXT}}

\def\secnuml#1\secnumr{\texorpdfstring{$#1$}{TEXT}}

\def\eqa{\begin{eqnarray}}
\def\eqae{\end{eqnarray}}
\def\eq{\begin{equation}}
\def\eqe{\end{equation}}
\def\be{\begin{equation}}
\def\ee{\end{equation}}
\def\bea{\begin{eqnarray}}
\def\eea{\end{eqnarray}}
\def\ba{\begin{array}}
\def\ea{\end{array}}
\def\bd{\begin{displaymath}}
\def\ed{\end{displaymath}}

\def\Tr{{\rm Tr}}

\def\>{\rangle}
\def\<{\langle}
\def\a{\alpha}
\def\b{\beta}
\def\c{\chi}
\def\del{\delta}

\def\f{\phi}

\def\g{\gamma}
\def\h{\eta}

\def\l{\lambda}

\def\w{\omega}
\def\p{\pi}

\def\r{\rho}
\def\s{\sigma}
\def\t{\tau}

\def\x{\xi}

\def\F{\Phi}
\def\G{\Gamma}

\def\S{\Sigma}

\def\pa{\partial}

\title{Vector models and generalized SYK models}
\author{Cheng Peng}
\affiliation{Department of Physics, Brown University, Providence RI 02912, USA}
\emailAdd{cheng$\underline{~}$peng@brown.edu}

\abstract{
We consider the relation between SYK-like models and vector models by studying a toy  model where a tensor field is coupled with a vector field.  By integrating out the tensor field, the toy model reduces to the Gross-Neveu model in 1 dimension. On the other hand, a certain perturbation can be turned on and the toy model flows to an SYK-like model at low energy. A chaotic-nonchaotic phase transition occurs as the sign of the perturbation is altered. We further study similar models that possess chaos and enhanced reparameterization symmetries.
}

\begin{document}

\maketitle

%%%%%%%%%%%%%%%%%%%%%%%%%%
\section{Introduction}
%%%%%%%%%%%%%%%%%%%%%%%%%%

The Sachdev-Ye-Kitaev (SYK) model~\cite{Sachdev:1992fk,PG,GPS,KitaevTalk1,KitaevTalk2,Maldacena:2016hyu}
is a quantum mechanical model that is  solvable in the  large-$N$ limit. At low energy an approximate reparametrization symmetry emerges and the dynamics of the model develops quantum chaos. These relevant features in black holes physics~\cite{Shenker:2013pqa,Shenker:2014cwa,Maldacena:2015waa,Sachdev:2015efa,Sachdev:2010um,Maldacena:2016hyu} make the SYK model promising to probe quantum gravity in AdS$_2$~\cite{Strominger:1998yg,Maldacena:1998uz,Almheiri:2014cka}.  Various aspects of the model have been studied extensively, such as
its spectral properties and its relation with the random matrix models~\cite{Polchinski:2016xgd,Jevicki:2016bwu,Maldacena:2016hyu,Jevicki:2016ito,Gross:2017hcz, Garcia-Garcia:2016mno, Garcia-Garcia:2017pzl, Cotler:2016fpe,Liu:2016rdi,Stanford:2017thb,Li:2017hdt}, its enhanced reparametrization symmetry~\cite{Maldacena:2016hyu,Bagrets:2016cdf,Stanford:2017thb}, the dilaton gravity theory with the same reparametrization symmetry~\cite{Maldacena:2016hyu,Maldacena:2016upp,Engelsoy:2016xyb,Grumiller:2016dbn,Cvetic:2016eiv,Forste:2017kwy} and its applications in condensed matter physics~\cite{Danshita:2016xbo,Garcia-Alvarez:2016wem,
DFGGJS,You:2016ldz,Fu:2016yrv,Hartnoll:2016apf}.
In addition, various extensions of this model have been considered, including models with supersymmetry~\cite{Fu:2016vas,Anninos:2016szt,Sannomiya:2016mnj}, models without quenched  disorder~\cite{Witten:2016iux,Gurau:2011xq,Bonzom:2011zz, Bonzom:2012hw,Gurau:2016lzk,KT,Carrozza:2015adg,NT,Peng:2016mxj, Krishnan:2016bvg,Krishnan:2017ztz, Gurau:2017xhf,Bonzom:2017pqs}, models in higher dimensions~\cite{Gu:2016oyy,Berkooz:2016cvq,Jian:2017unn,Turiaci:2017zwd,Gu:2017ohj}, models with extra symmetries~\cite{Gross:2016kjj}, and models with phase transitions~\cite{Banerjee:2016ncu,Bi:2017yvx,Jian:2017jfl}. Other related works can be found in~\cite{Jensen:2016pah,Ferrari:2017ryl, Berkooz:2017efq,Bagrets:2017pwq,Caputa:2017urj,Chowdhury:2017jzb,Cotler:2017abq,Itoyama:2017emp,Ho:2017nyc}.

The SYK model also bears some resemblance with large-$N$ vector models.
The SYK model possesses a tower of composite operators, which is similar to the tower of higher spin conserved currents in the $O(N)/U(N)$ vector models.  The SYK model is strongly coupled, so the operators therein acquire $\co(1)$ anomalous dimensions in the IR. On the contrary, the operators in the  vector models acquire small anomalous dimensions that are suppressed by $\frac{1}{N}$. It is widely believed that
the different weakly coupled vector models are holographically dual to higher-spin theories~\cite{Klebanov:2002ja,Sezgin:2003pt, Giombi:2009wh, Koch:2010cy, Gaberdiel:2010pz,Giombi:2012ms, Giombi:2011kc,Maldacena:2011jn, Maldacena:2012sf,Vasiliev:1995dn}, which could be regarded as suitable tensionless limits of String theory~\cite{Bianchi:2003wx,Chang:2012kt, Aharony:2013ipa, Gaberdiel:2014cha,Gaberdiel:2015uca}. On the other hand, the SYK model is believed to be dual to a stringy theory that effectively has finite string  tension~\cite{Maldacena:2016hyu}. As a result,  it would be illuminating to understand the relation between SYK-like models and vector models, which could shed some light on
the effect of changing the string tension in the dual theory.

In this work, we use a toy model \eqref{action} to probe the connection between SYK-like  models and vector models. We show that the model \eqref{action} is nonchaotic and reduces to the Gross-Neveu model by integrating out the tensor field. We then turn on a perturbation at the infrared SYK$_2$-like fixed point of \eqref{action}. The perturbation involves an additional bosonic vector field. The nature of this perturbation depends on its sign. If the sign is negative, the perturbation is marginally irrelevant and the theory remains nonchaotic. If the sign is positive, the perturbation becomes marginally relevant and the model flows to a different model \eqref{H1}. At low energy, this new model is chaotic and possesses enhanced reparameterization symmetry.
Therefore changing the sign of the perturbation across zero leads to a transition between a chaotic phase and a nonchaotic phase, which is similar to those observed in the models with quenched disorder~\cite{Bi:2017yvx, Jian:2017jfl}.
This sets up a connection between the Gross-Neveu vector model and an SYK-like chaotic model.
 We further study similar models that couple tensor fields with vector fields, whose IR dynamics are slight different from \eqref{H1}. Nevertheless, all of them are shown to be chaotic.

%%%%%%%%%%%%%%
\section{A model with a charge-charge interaction}
\label{minimal}

We start with a model with a simple coupling. This model can be regarded as a tensor generalization of earlier toy matrix models~\cite{Festuccia:2006sa,Iizuka:2008eb} that are used to explore black hole thermalization and the information paradox from the field theory perspective.

%%%%%%%%%
\subsection{The model}\label{mainmodel}

We consider a quantum mechanical model of a complex fermionic tensor $\l^{abc}$ and a complex fermionic vector $ {\c}^a$
\bal
H_0&=\frac{J}{2\,N^{3/2}}\,\Big( \bar{\l}^{ab}{}_{i} \bar{\l}^{ab}{}_{{j}}{\c}^i {\c}^j +{\l}^{abi} {\l}^{abj}\bar{\c}_i \bar{\c}_j \Big)\,, \qquad J\sim \co(N^0) \ . \label{action}
\eal
The model possesses an $O(N)\times O(N) \times U(N)$  symmetry.
The indices of the tensor field are distinguished by their positions; the first two indices are vector indices of the first two $O(N)$ factors respectively, and the last upper (lower) index labels the fundamental (anti-fundamental) representation of the $U(N)$ group.
The complex fermion
transforms in the fundamental representation of the $U(N)$ factor, and is a  singlet of the  $O(N)$ factors.

\begin{figure}[t]
  \begin{center}
\includegraphics[width=0.20\linewidth]{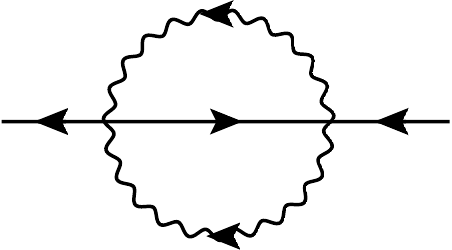}
  \end{center}
\caption{The fundamental melon that dominates all the corrections to the two point function of the $\c^i$  field. The wavy lines in the diagram represent tensor field,
and the solid lines represent the vector fermion. Direction of the arrows distinguishes a field from its conjugation; any line with its arrow going into (out of) a vertex represents a (conjugate) field.}
\label{fig:melon1}
\end{figure}

We now solve this model in the large-$N$ limit with fixed $J$.
It is easy to check that loop corrections to the two point function of the tensor field are
all suppressed by powers of $1/N$. Therefore the $\l^{abi}$ propagator remains classical at the leading order of~$N$
\bal
G^\l(\t_1,\t_2)=\frac{1}{2}\text{sgn}(\t_{12})\,,  \label{clal}
\eal
where we adopt the short-hand notation $\t_{12}\equiv\t_1-\t_2$.
 On the other hand, the two point function of the vector field receives corrections from the iterative insertions of the melon depicted in figure~\ref{fig:melon1}.
The Schwinger-Dyson equations for the 2 point function of the vector field are
  \bal
  &\S^\c(\t_1,\t_2)=  \frac{J^2}{4} G^\l(\t_1,\t_2)^2\, G^{\c}(\t_1,\t_2)
  \label{eq1}\\
  &\pa_\t G^\c(\t_1,\t_3)+\int d\t_2 G^\c(\t_1,\t_2)\S^\c(\t_2,\t_3)=-\delta(\t_{13})\ .\label{eq2}
  \eal
  In the low energy/strong coupling regime, the kinetic term
  is sub-dominant and the equations
 can be solved by
 \bal
 G^\c_c(\t_1,\t_2)&=\frac{2\,\text{sgn}(\t_{12})}{J \p |\t_{12}|}\ .\label{ccir}
 \eal
 It is easy to verify that the set of equations \eqref{eq1} and \eqref{eq2} takes the same form as the $q=2$ SYK model (SYK$_2$) once  \eqref{clal} is plugged in, so does the solution \eqref{ccir}.

\begin{figure}[t]
  \begin{center}
\includegraphics[width=0.350\linewidth]{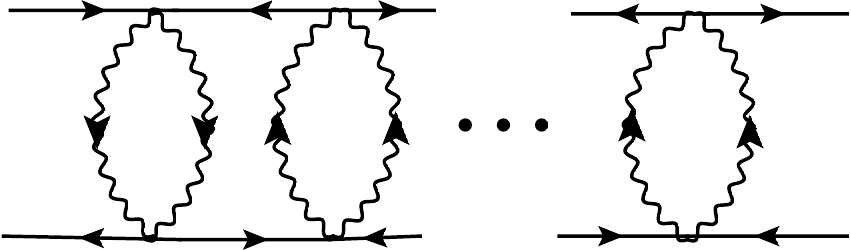}
  \end{center}
\caption{The ladder diagrams that dominate  the 4-point correction functions of the vector fields. The solid and wavy lines represent the vector and tensor field respectively. }
\label{fig:ladder1}
\end{figure}

 The similarity with the SYK$_2$ model continues to the 4-point function.  The connected piece of the $\<\bar\c_i(\t_1){\c}^i(\t_2)\bar{\c}_j(\t_3)\c^j(\t_4)\>$ 4-point function is dominated by the ladder diagrams   shown in figure~\ref{fig:ladder1}.
The kernel of such ladder diagrams is
\bal
K(\t_1,\t_2;\t_3,\t_4)&=\frac{J^2}{4}  \, G^{\c}(\t_1,\t_4)\, G^{\c}(\t_2,\t_3)G^\l(\t_3,\t_4)^2\ .
\eal
We would like to learn if this model is chaotic, which can be diagnosed by the out-of-time-order correlation function~\cite{Maldacena:2016hyu}
						\bal
						\Tr\left(y \bar\c_{{i}}(t_1) y \c^{{j}}( 0) y \c^{{i}}(t_2) y \bar\c_{{j}}( 0) \right)\,, \qquad y=\r(\b)^{\frac{1}{4}}\,,\label{qtt1}
						\eal
						where $\r(\b)$ is the thermal partition function. Equation \eqref{qtt1}  is closely related to the quantity
						\bal
						\<y^2 [\bar\c_{{i}}(t ), \c^{{j}}( 0)] y^2 [\bar\c_{{i}}(t ), \c^{{j}}( 0)]^\dagger\>\,,
						\eal
						that diagnoses quantum chaos at late time~\cite{Shenker:2013pqa, Maldacena:2015waa}. This quantity fixes the integration contour and hence the way we analytically continue the kernels onto the contour with two real time folds~\cite{ KitaevTalk1,Maldacena:2016hyu}.

There are two subtleties of this continuation. Firstly, whenever a vertex is  inserted on the real time fold, we need to include a factor of $i$. This comes from the continuation of the vertex from Euclidean to Lorentzian signature. Secondly, depending on the form of the interaction, each of the two fields coming out of the kernel might not be on the same real time fold as the corresponding incoming field if we were doing the continuation naively. The correct continuation that gives the kernel for the quantity \eqref{qtt1} is to first switch the two fields that would be on the thermal circle (this is equivalent to rotate both of them by half of the circle)
before the continuation. Notice that this might give an extra sign if the two fields we rotated are both fermions.

We now determine the retarded kernel in our model with these two factors taken into account. Firstly, there is a factor of $i^2$ due to the two vertices inserted on the real time folds. Secondly, our kernel does have the ``wrong" order as discussed above.
This can be verified by the directions of the arrows on the propagators on the sides of the ladder. A simpler way to determine this is  from the fact that the kernel contains the propagators $G(\t_{1},\t_4) $ and $G(\t_{2},\t_3)$ on the sides, rather than the ones with the correct order, namely $G(\t_{1},\t_3)$ and $ G(\t_{2},\t_4)$.   So as discussed above, we have to switch the operators on the thermal circle. This results in switching the label 3 and 4 in the expression together with an extra minus sign. Finally,
we replace the propagators on the sides by the corresponding retarded propagators and  replace the rest propagators by the ``ladder rung" propagators.
This leads to the following retarded kernel						\bal
K_R(\t_1,\t_2;\t_3,\t_4)&= \frac{J^2}{4}  \, G_R^{\c}(\t_1,\t_3)\, G_R^{\c}(\t_2,\t_4)G^\l_{lr}(\t_3,\t_4)^2\ .
\eal
We then observe that this is the same as the retarded kernel of the SYK$_2$ model~\cite{Maldacena:2016hyu},
which
is not chaotic~\cite{Maldacena:2016hyu,Gross:2016kjj}. This is not a surprise since the very similar IOP model is not chaotic as well~\cite{Michel:2016kwn}.

Notice that the $O(N)\times O(N)\times U(N)$ symmetry is not crucial for the construction: we can use an $U(N)\times O(N)\times U(N)$ symmetric tensor to construct a similar model as \eqref{action}
but then we have to include $2\times 2=4$ different fields, that are in the \mbox{(anti-)fundamental} of the first and last $U(N)$ factor respectively.  We will consider tensor fields with the $U(N)\times O(N)\times U(N)$ symmetry for slightly different models in section \ref{uou}.

Up to now, the model does not seem very exciting. However, as we will show in the following sections, it has close relations with both the 1-dimensional Gross-Neveu  vector model and many SYK-like solvable and chaotic  models.

 %%%%%%%%%%%%%%%%%%%%%%
 \subsection{Relation to the Gross-Neveu model}

 Because the tensor field is complex, we can introduce a mass term  $M\,  \bar{\l}^{ab}{}_{{i}}\l^{abi} $.
 as long as $ M\ll J$, the discussion in section~\ref{mainmodel} remains valid.

\begin{figure}[t]
\begin{center}
\includegraphics[width=0.50\linewidth]{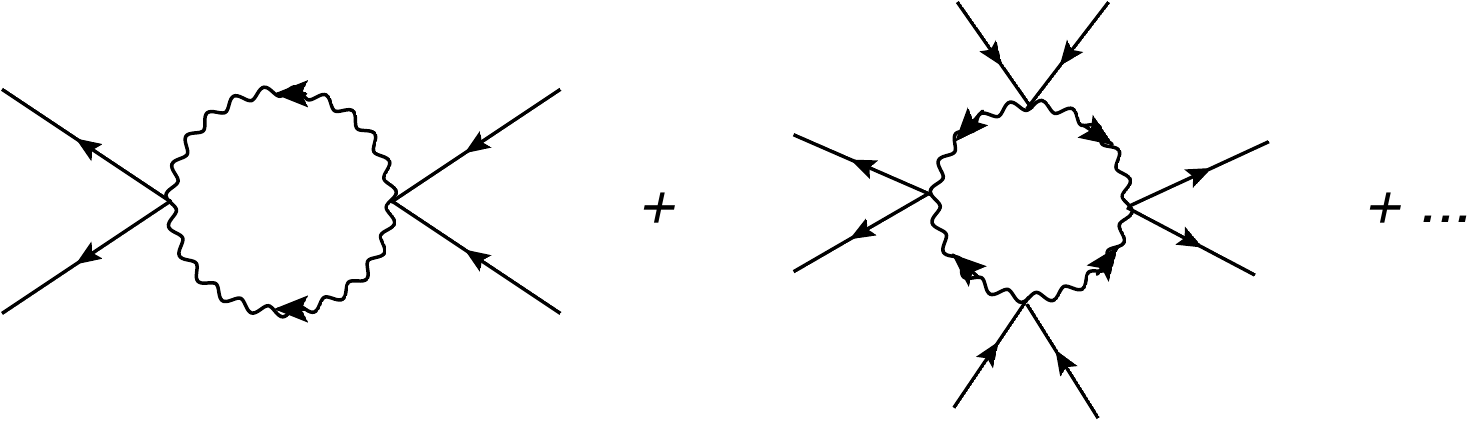}
\end{center}
\caption{The effective coupling of the vector model obtained by integrating out the tensor fields. Notice that a similar 6-point effective coupling is not generated due to the form of the interaction \eqref{action}.}
\label{fig:efv}
\end{figure}

On the other hand, if the tensor field is heavy enough, namely $J\ll M$, we can integrate out the tensor field  to get a vector model.
In practice, this amounts to compute the 1PI effective action of the vector field  by summing over all tensor loops, which generates a tower of $2k$-point interactions
$\co_k\sim (\bar\c_i\c^i)^k$, where $k$ is even.
Diagrammatically, this is
illustrated  in figure~\ref{fig:efv}.
For example, the quartic interaction of the vector model can be obtained by integrating out the tensor fields in the 4-point function shown in the first panel of figure~\ref{fig:efv}
\bal
O_2(\t) &= \int d\t' \frac{J^2}{4 N^3} N^2 (G^\l(\t-\t'))^2 \c^i(\t)\c^j(\t)\bar\c_j(\t')\bar\c_i(\t')\ .\label{toGN}
\eal
The propagator of the tensor field is
\bal
&G^\l(\w)=-\frac{1}{i\w+M} \\
&G^\l(\t,\t')=\frac{1}{2\p}\int d\w e^{-i \w (\t-\t')}G^\l(\w)= -  e^{M (\t-\t')} \theta (\t'-\t) \,,\label{gltt}
\eal
where $\theta$ is the unit step function.
The integration \eqref{toGN} can be carried out straightforwardly and gives
\bal
O_2(\t)
&=  \frac{J^2}{8 N M}  \left( \c^i(\t)\c^j(\t)\bar\c_j(\t)\bar\c_i(\t)+  \frac{1}{M}\c^i(\t)\c^j(\t)\bar\c_j(\t)\pa_\t \bar\c_i(\t)+\ldots\right)\ .
\eal
To the leading order in $\frac{1}{M}$, we get
\bal
O_2(\t) = \frac{J^2}{8 N  M}
(\bar\c_i\c^i)^2=\frac{g}{N} (\bar\c_i\c^i)^2\,, \qquad g=\frac{J^2}{8 M}\sim \co(N^0)
\,,\label{OGN2}
\eal
upto other term surpressed by $\frac{1}{N}$ and/or  $\frac{1}{M}$. Higher point vertices in the vector model can be worked out similarly, in particular the $N$ dependence of the $2k$ point function is $\co_k\sim N^{2-\frac{3}{2}k}(\bar\c_i\c^i)^k$. Such $N$ dependence indicates that all other effective vertices with $k>2$
do not contribute to the correlation functions at the leading order of~$N$;~contributions with such $\co_k$ insertions are always suppressed by $\frac{1}{N}$ comparing to the contributions with only $\co_2$ insertions.
Therefore the resulting vector model effectively contains only a quartic interaction \eqref{OGN2} at the leading order of $N$ and $M$,
which is nothing but the interaction of the Gross-Neveu model in 1 dimension.

\begin{figure}[t]
\begin{center}
\includegraphics[width=0.80\linewidth]{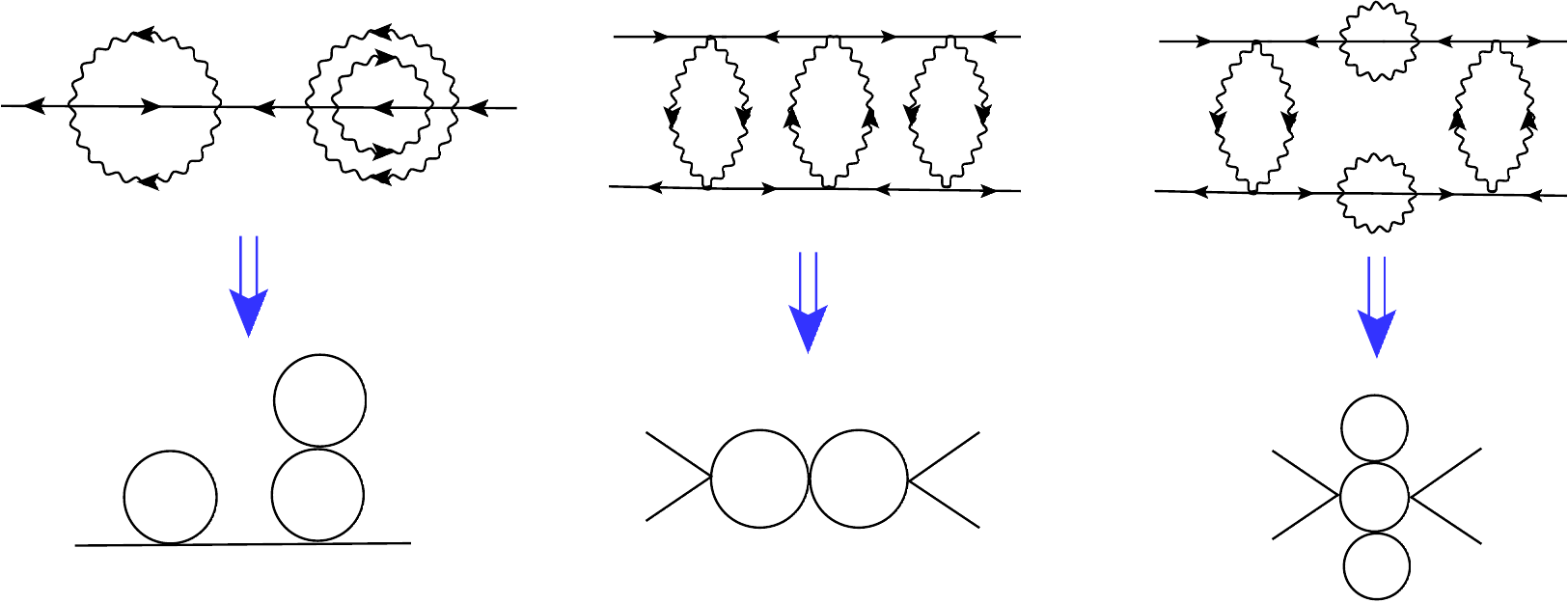}
\end{center}
\caption{Mapping the melonic diagrams of the model \eqref{action} onto the bubble diagrams of the vector model in the limit $J\ll M$. Here we have assumed a (small) mass term for the vector fermion. If this is not the case, the snail diagrams in the vector model and the corresponding diagrams in the model \eqref{action}  vanish.}
\label{fig:maptoGN}
\end{figure}

In fact, integrating out the tensor fields in all the melonic diagrams of model \eqref{action} gives the bubble diagrams, for example those in the second line of figure~\ref{fig:maptoGN}, that indeed dominate the large-$N$ limit of the Gross-Neveu model.
This confirms our assertion that only the 4-point vertex is relevant in the large-$N$, large-$M$ limit of the  corresponding vector model.
For example, starting from the 1-melon diagram of the $\<\bar\c_i(\t_1)\c^i(\t_2)\>$ 2-point function, integrating out the two tensor propagators gives the following nonvanishing contribution to the snail diagram of the resulting vector model
\bal
\s_{t_{12}}&= \int d\t d\t' \frac{J^2}{4 N^3} N^3 (G^\l(\t-\t'))^2 G^\c(\t_1-\t)G^\c(\t-\t')G^\c(\t'-\t_2)\\
&\cong
\int_{-\infty}^{\infty} d\t \frac{J^2}{4  }   G^\c(\t_1-\t)G^\c(0)G^\c(\t-\t_2) \frac{1}{2M}+\co(M^{-2})\,,
\eal
where in the last step we keep only the leading term in the limit $\w\ll M$. This can also be verified by a simple saddle point argument. Then it is clear to see that the prescribed dictionary is consistent: when $G^\c(0)=0$, both the snail diagram in the vector model and the corresponding diagram in the tensor model vanish.

As another consistent check, we can integrate out the tensor fields in the Schwinger-Dyson equation \eqref{eq2} of the model \eqref{action}. This is different from what we did in section \ref{mainmodel} since we are now in a different regime $M\gg \w$. The result in this limit is
\bal
  &\pa_{\t_{13}} G^\c(\t_{13})+\int d\t_2 G^\c(\t_1,\t_2) \frac{J^2}{4}  G^\l(\t_2,\t_3)^2\, G^{\c}(\t_2,\t_3)=-\delta(\t_1-\t_3)\,,
\eal
where we have explicitly used the fact that $G^\c(\t_1,\t_2)=G^c(\t_{12})$.
Plugging in \eqref{gltt} and using the saddle point analysis again, the integral localizes at $\t_2=\t_3$, and the result is simply
\bal
&\left(\pa_{\t_{13}}+\frac{J^2}{8M}  \, G^{\c}(0)\right) G^\c(\t_{13}) =-\delta(\t_{13})\ .\label{tocmp}
\eal
On the other hand, the gap equation of the fermion bilinears $\s(\t)=\frac{g}{N}\bar\c_i(\t)\c^i(\t)$ in the resulting vector model is
\bal
\frac{\s}{g}=\frac{\del}{\del\s}\left(\ln\det(-\pa_\t-\s)\right)\,,\label{gap}
\eal
We observe that \eqref{gap} can be, roughly speaking, regarded as the singular $\t_{13}\to 0$ limit of \eqref{tocmp} once we carry out the variation on the functional determinant and notice that
\bal
\s=\frac{g}{N}\<\bar\c_i(\t)\c^i(\t)\>=g \, G^\c(0)\ .
\eal
This is compatible with our connection between the model \eqref{action} and the Gross-Neveu model. It will be useful to clarify this identification of  \eqref{gap} and \eqref{tocmp} with a more careful treatment of the functional determinant and its variation.

The discussion in this subsection is generic. It applies equally well to higher dimensional analogues of our model \eqref{action}, although a more sophisticated procedure of integrating out the tensor field should be used.
In addition, this discussion generalizes straightforwardly to relations between the bosonic tensor models and the bosonic $O(N)/U(N)$ vector models. Moreover,  the fact that the first factor of the symmetry group being $O(N)$ instead of $U(N)$ is not crucial in the discussion; the result extends trivially to similar models  of uncolored tensor models with $U(N)\times O(N)\times U(N)$ symmetry.

%%%%%%%%%%%%
\subsection{A phase transition to a chaotic model}\label{pert0}
%%%%%%%%%%
As discussed in many other SYK-like models, the model \eqref{action} could flow to other models if we turn on some perturbations. For example, consider the following perturbed model
\bal
H'_0&=\frac{J}{2\,N^{3/2}}\,\Big( \bar{\l}^{ab}{}_{i} \bar{\l}^{ab}{}_{{j}}{\c}^i {\c}^j +{\l}^{abi} {\l}^{abj}\bar{\c}_i \bar{\c}_j \Big)+\frac{u}{N^2} \bar\l^{ab}{}_i\l^{ebj}\f^a\f^e\c^i\bar\c_j\ .\label{dh2}
\eal
where the coupling $u$ is infinitesimal and of order $N^0$. The $\f^b$ is a free real vector boson transforming in the vector representation of the first $O(N)$ factor.

The reason to turn on this perturbation can be understood as follows. We start with a model with a free fermionic tensor field $\l^{abi}$,  a free fermionic vector field $\c^i$ and a free bosonic vector field $\f^a$. We can first turn on a coupling between the tensor $\l^{abi}$ and the fermion  $\c^i$, whose simplest form is \eqref{action}. As we have discussed in section \ref{mainmodel}, this model is nonchaotic. Since a similar minimal coupling between the tensor $\l^{abi}$ and the boson $\f^a$ is forbidden by the Fermi statistics, the next simplest coupling involving the tensor field is this perturbation in \eqref{dh2}. There is another type of deformation involving the self-interactions of the tensor field. This is a different type of deformation since it will make the tensor field dominate the vector fields in a brute-force way, which we will discuss more in section \ref{tself}.

The running of the coupling $u$ is again dominated by the melonic diagrams, but now we do not need to sum all the contribution: the diagram with two $u$ vertices insertion is enough.
The correction to the vertex \eqref{dh2} is
\bal
\delta u&=2\int d\t_2 u^2 G^\l(\t_1,\t_2) G^\f(\t_1,\t_2)  G^\c(\t_1,\t_2)\ .
\eal
Plugging in the IR solution \eqref{ccir} of the model \eqref{action}, together with the propagator of the free boson $G^\f(\t_1,\t_2)=1$, it becomes
\bal
\delta u&=\frac{2u^2}{ J\p }\int d\t_2    \frac{ 1}{  |\t_{12}|}
= \frac{4 u^2}{ J\p } \ln\left(\frac{L}{\ell}\right)\,,
\eal
 where  $L$ is a cutoff scale and $\ell$ is the renormalization scale.
The RG running of the coupling is then the running of $u$ as a function of $1/\ell$. The contribution from the wavefunction renormalization is at higher order in $u$, we therefore have
 \bal
 \b(u)\equiv\frac{du}{d\ln(1/\ell)}= - \frac{4 }{ J\p } u^2\ .
 \eal
When the perturbation $u<0$, it is marginally irrelavent and the SYK$_2$ fixed point \eqref{ccir}  of the model \eqref{action} is stable. If the initial perturbation $u>0$, it is marginally relavent and the
theory flows to another fixed point that is dominated by the interaction
\bal
 H_1&=\frac{k}{N^2} \bar\l^{ab}{}_i\l^{ebj}\f^a\f^e\c^i\bar\c_j\ . \label{H0p}
\eal
This is a similar phase transition to what is observed in~\cite{Bi:2017yvx, Jian:2017jfl}. As discussed there, this is a phase transition like the Kosterlitz-Thouless transition~\cite{Kosterlitz:1973xp}  in the sense that the critical exponent of the dynamical energy scale of the system diverges as the sign of the perturbation is changed across zero.

We study  the new IR fixed point of the model defined by \eqref{H0p}  in the next section. We  will demonstrate that the  inclusion of the ``auxiliary" vector boson does modify the IR property of the model significantly and
 the new fixed point is chaotic.

%%%%%%%%%%%%%%%%%%
\section{A chaotic model with a tensor-vector coupling
} \label{tffsec}

As discussed in the previous section, the model \eqref{action}  flows to a model described by the Hamiltonian \eqref{H0p} after a phase transition. In this section we study the physics at the low energy/strong coupling limit of the model \eqref{H0p}, which is a natural fixed point of the flow discussed in section \ref{pert0}.

\subsection{The model}

We study the model \eqref{H0p}, which we recast here
\bal
 H_1&=\frac{k}{N^2} \bar\l^{ab}{}_i\l^{ebj}\f^a\f^e\c^i\bar\c_j\ . \label{H1}
\eal
We emphasis again that the vector boson and the vector fermion are charged under different symmetry groups.

The corrections to the $\l$ field  are subleading in $1/N$,
so in the $M\ll k,g$ limit the $\l$ field remains free
\bal
G^\l(\t_1,\t_2)&=\frac{1}{2}\text{sgn}(\t_{12})\ .\label{freel}
\eal
In the low energy limit, the large-$N$ Schwinger-Dyson equations for the $\c$ and the $\f$ fields are
\bal
&\S^\c(\t_1,\t_2)=  k^2 G^\l(\t_1,\t_2)^2\, G^{\c}(\t_1,\t_2)\, (G^{\f}(\t_1,\t_2))^2
\label{sd1}\\
&\S^\f(\t_1,\t_2)= 2k^2 G^\l(\t_1,\t_2)^2\, (G^{\c}(\t_1,\t_2))^2\,G^{\f}(\t_1,\t_2)
\\
& \int d\t_2 G^\c(\t_1,\t_2)\S^\c(\t_2,\t_3)=-\delta(\t_{13})\\
& \int d\t_2 G^\f(\t_1,\t_2)\S^\f(\t_2,\t_3)=-\delta(\t_{13})\ .\label{sd2}
\eal
The factor of 2 in the $\S^\f(\t_1,\t_2)$ equation comes from the two different flows of the arrow in the melon; both are allowed since the bosons are real.
We take an ansatz
\bal
G^\c(\t_1,\t_2)=\frac{b^\c\text{sgn}(\t_{12})}{|\t_{12}|^{2\Delta_\c}}\,, \qquad
G^\f(\t_1,\t_2)=\frac{b^\f }{|\t_{12}|^{2\Delta_\f}}\,, \label{cfir}
\eal
for the vector fields and the Schwinger-Dyson equations become
\bal
\frac{1}{4} k^2(b^\f)^2 (b^\c)^2 & \int d\t_2 \frac{\text{sgn}(\t_{12})}{|\t_{12}|^{2\Delta_\c}} \frac{ \text{sgn}(\t_{23})}{|\t_{23}|^{2\Delta_\c+4\Delta_\f}}
=-\delta(\t_{13})\\
\frac{1}{2} k^2 (b^\c)^2 (b^\f)^2& \int d\t_2 \frac{1 }{|\t_{12}|^{2\Delta_\f}}  \, \frac{1 }{|\t_{23}|^{2\Delta_\f+4\Delta_\c}} =-\delta(\t_{13})\ .
\eal
Using the integrals
 \bal
&\int_{-\infty}^{\infty}   d x  \frac{\text{sgn}(x-a)\text{sgn}(x-b)}{|x-a|^{2\a}  |x-b|^{2\b} }=\begin{cases}
\p   \frac{\cos(\p(\a+\b-1/2))\G(2\a+2\b-1)}{  \sin(\p\a)\Gamma(2\a) \sin(\p\b)\G(2\b)} \frac{1}{|a-b|^{2\a+2\b-1} }\qquad \a+\b\neq 1\\
   \frac{\p^2}{ \sin(\p\a)\Gamma(2\a) \sin(\p\b)\G(2\b)}\delta(a-  b)  \,, \qquad \a+\b= 1
\end{cases}\,, \label{todelta}
\eal
\bal
&\int_{-\infty}^{\infty}   d x  \frac{1}{|x-a|^{2\a}  |x-b|^{2\b} }
=
\begin{cases}
	\p   \frac{\cos(\p(\a+\b-1/2))\G(2\a+2\b-1)}{ \cos(\p\a)\Gamma(2\a) \cos(\p\b)\G(2\b)} \frac{1}{|a-b|^{2\a+2\b-1} }\,,\qquad \a+\b\neq 1\\
	\frac{\p^2}{ \cos(\p\a)\Gamma(2\a)  \cos(\p\b)\G(2\b)}\delta (a-  b)\,,\qquad \a+\b= 1
\end{cases}\,,\label{todeltas}
\eal
the above equations have solutions when
\bal
\Delta_\c+\Delta_\f=1/2\ .\label{rel1}
\eal
Plugging in this condition, the integrals reduce to
\bal
&   \frac{(\p)^2 \frac{1}{4} k^2(b^\f)^2 (b^\c)^2}{ \sin(\p\Delta_\c)\Gamma(2\Delta_\c) \sin(\p(1-\Delta_\c))\G(2-2\Delta_\c )}  =1\\
& \frac{\p^2\frac{1}{2} k^2 (b^\c)^2 (b^\f)^2}{ \cos(\p(1/2-\Delta_\c))\Gamma(1-2\Delta_\c)  \cos(\p(1/2+\Delta_\c))\G(1+2\Delta_\c)}=-1\ .
\eal
They are uniquely solved by
\bal
\Delta_\c=\frac{1}{3}\,, \qquad \Delta_\f=\frac{1}{6}\,,\qquad (b^\f)^2 (b^\c)^2=\frac{2\sqrt{3}}{3 \pi  k^2}\ .
\eal

\begin{figure}[t]
\begin{center}
\includegraphics[width=0.90\linewidth]{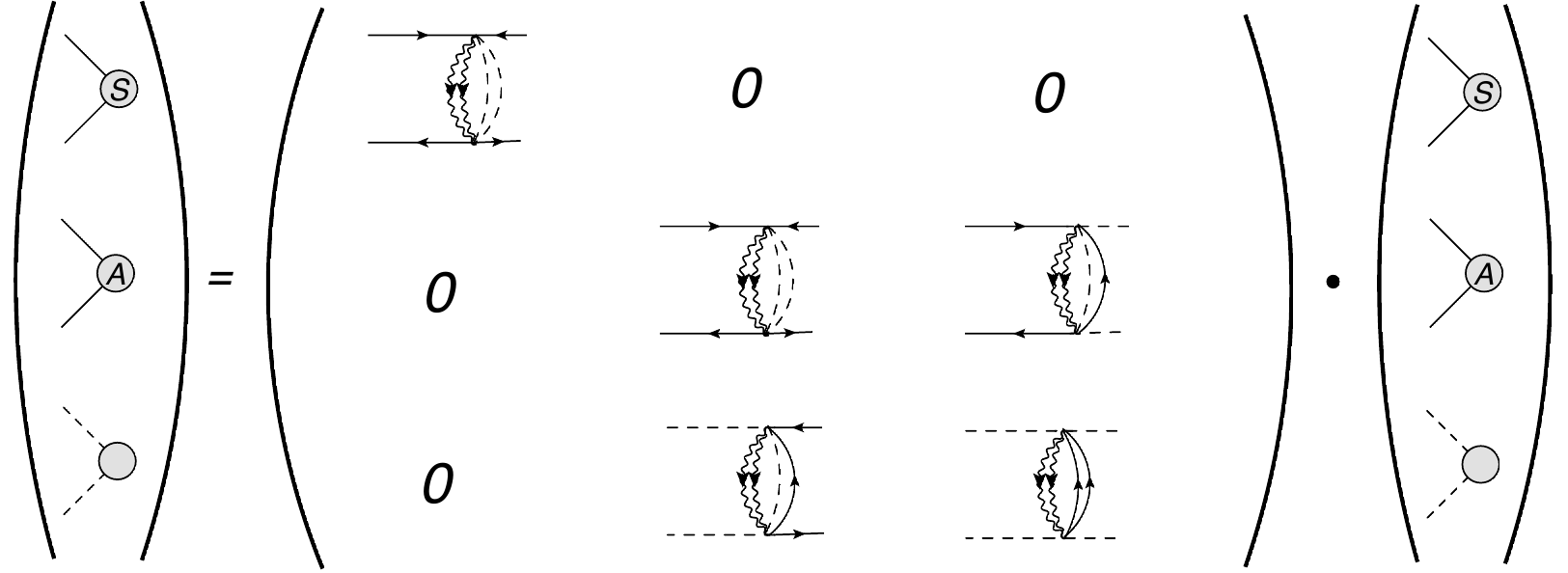}
\end{center}
\caption{Schematic form of the eigen-equations associated to the kernels of the 4-point functions in the model \eqref{H1}. The solid, dashed and wavy lines represent the vector fermions, the vector bosons and the tensor fields respectively. The letters in the blobs of the eigenvectors indicate  whether it is the symmetric or antisymmetric eigenfunction.}
\label{fig:ska}
\end{figure}

		The 4-point function of the tensor fields is again not corrected at the leading order in $\frac{1}{N}$. The 4-point function of the vector fermions $\<\bar\c_i(\t_1)\c^i(\t_2)\bar\c_j(\t_3)\c^j(\t_4)\>$
		 mixes with the other two correlators $\<\bar\c_i(\t_1)\c^i(\t_2)\f^a(\t_3)\f^a(\t_4)\>$ and $\<\f^a(\t_1)\f^a(\t_2)\f^b(\t_3)\f^b(\t_4)\>$. The mixing is conducted by 4 different kernels
		\bal
		K^{(11)}&=-k^2 G^\c(\t_1,\t_3)G^{\c}(\t_2,\t_4) (G^\l(\t_4,\t_3))^2 (G^\f(\t_4,\t_3))^2\\
		K^{(12)}&=-2 k^2 G^\c(\t_1,\t_3)G^{ \c}(\t_2,\t_4) (G^\l(\t_3,\t_4))^2 G^\f(\t_3,\t_4)G^\c(\t_3,\t_4) \\
		K^{(21)}&=4 k^2 G^\f(\t_1,\t_3)G^{\f}(\t_2,\t_4) (G^\l(\t_3,\t_4))^2 G^\f(\t_3,\t_4)G^\c(\t_3,\t_4)\\
		K^{(22)}&=2k^2 G^\f(\t_1,\t_3)G^{\f}(\t_2,\t_4) (G^\l(\t_3,\t_4))^2 (G^\c(\t_3,\t_4))^2\ .
		\eal
		Since the boson $\f^a$ is real, there are 3 types of eigenvectors: the symmetric and antisymmetric fermionic $U^{s/a}$ together with a symmetric bosonic  $V$
		\bal
		U^s(\t_1,\t_2)&=\frac{1}{|\t_1-\t_2|^{2\Delta_\c-h}}=\frac{1}{|\t_1-\t_2|^{\frac{2}{3}-h}}\label{us}\\
		U^a(\t_1,\t_2)&=\frac{\text{sgn}(\t_1-\t_2)}{|\t_1-\t_2|^{2\Delta_\c-h}}=\frac{\text{sgn}(\t_1-\t_2)}{|\t_1-\t_2|^{\frac{2}{3}-h}}\label{us}\\
		V(\t_1,\t_2)&=\frac{1}{|\t_1-\t_2|^{2\Delta_\f-h}}=\frac{1}{|\t_1-\t_2|^{\frac{1}{3}-h}}\ .\label{V}
		\eal
		As a result, the kernel matrix to diagonalize is actually $3\times 3$, as shown in figure~\ref{fig:ska}.
		The eigenvalues of the kernels
		\bal
		&K^{(11)}U^s(\t_3,\t_4)=k^{11}U^s(\t_1,\t_2)\\
		&K^{(11)}U^a(\t_1,\t_2)=k^{22}U^a(\t_1,\t_2)\,,\quad K^{(12)}V(\t_3,\t_4)=k^{23}U^a(\t_1,\t_2)\\
		&K^{(21)}U^a(\t_3,\t_4)=k^{32}V(\t_1,\t_2)\,,\quad K^{(22)}V(\t_3,\t_4)=k^{33}V(\t_1,\t_2)\,,
		\eal
		can be evaluated to be
		\bal
		k^{11}&= \frac{2 \sqrt{3} \pi  \sin \left(\frac{1}{6} (3 \pi  h+\pi )\right) \sec \left(\frac{1}{6} \pi  (3 h+2)\right) \Gamma \left(\frac{2}{3}-h\right)}{\Gamma \left(-\frac{1}{3}\right)^2 \Gamma \left(\frac{4}{3}-h\right)}\\
		k^{22}&=-\frac{2 \sqrt{3} \pi  \cos \left(\frac{1}{6} (3 \pi  h+\pi )\right) \csc \left(\frac{1}{6} \pi  (3 h+2)\right) \Gamma \left(\frac{2}{3}-h\right)}{\Gamma \left(-\frac{1}{3}\right)^2 \Gamma \left(\frac{4}{3}-h\right)}\\
		k^{23}&=-\frac{4 \sqrt{3} \pi  \cos \left(\frac{1}{6} (3 \pi  h+\pi )\right) \csc \left(\frac{1}{6} \pi  (3 h+2)\right) \Gamma \left(\frac{2}{3}-h\right)}{\Gamma \left(-\frac{1}{3}\right)^2 \Gamma \left(\frac{4}{3}-h\right)}\\
		k^{32}&=-\frac{2 \sqrt{3} \pi  \sin \left(\frac{1}{6} \pi  (3 h+2)\right) \sec \left(\frac{1}{6} (3 \pi  h+\pi )\right) \Gamma \left(\frac{1}{3}-h\right)}{\Gamma \left(-\frac{2}{3}\right)^2 \Gamma \left(\frac{5}{3}-h\right)}\\
		k^{33}&=\frac{\sqrt{3} \pi  \sin \left(\frac{1}{6} \pi  (3 h+2)\right) \csc \left(\frac{1}{6} \pi  (8-3 h)\right) \Gamma \left(\frac{1}{3}-h\right)}{\Gamma \left(-\frac{2}{3}\right)^2 \Gamma \left(\frac{5}{3}-h\right)} \ .
		\eal

		\begin{figure}[t]
			\begin{center}
				\includegraphics[width=0.60\linewidth]{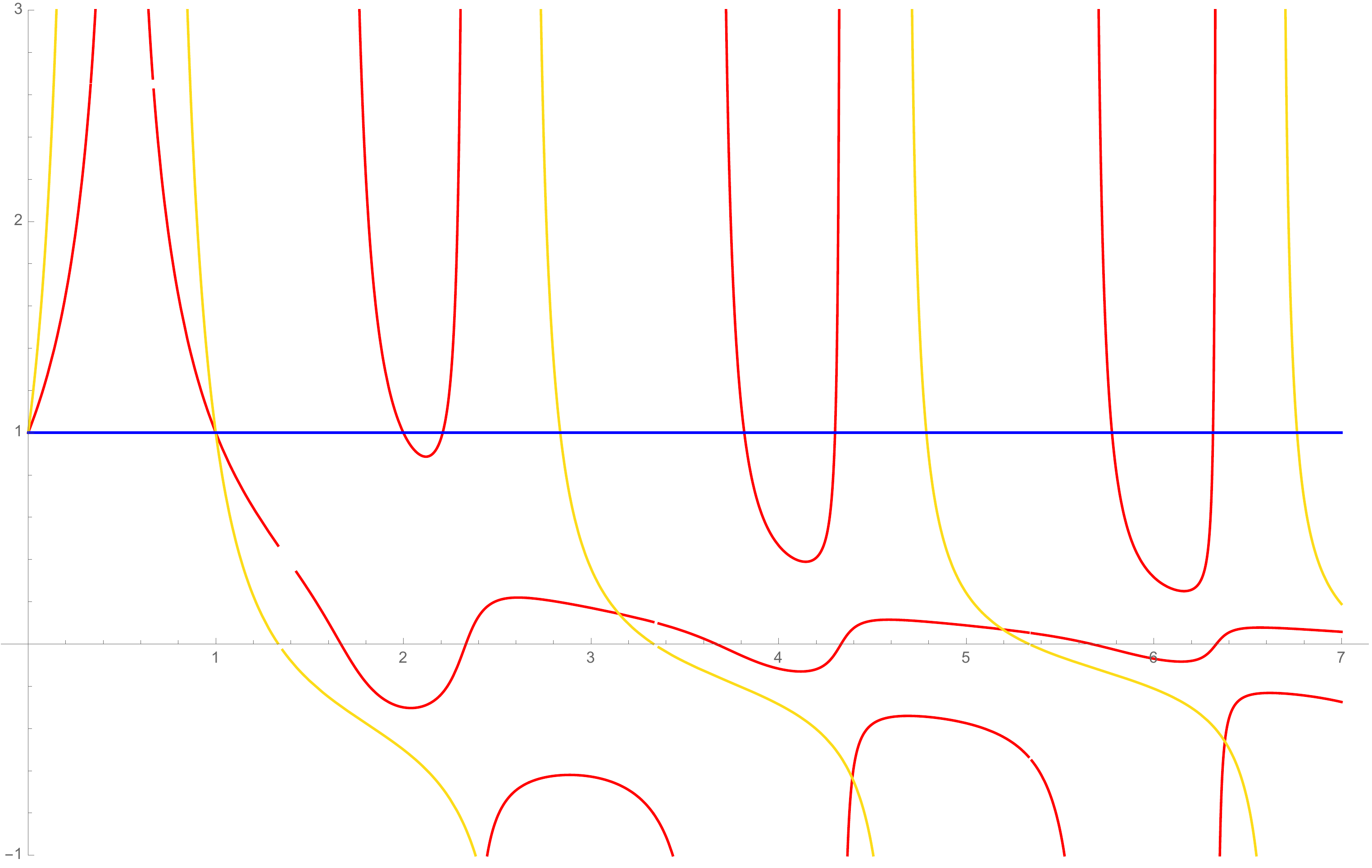}
				\end{center}
				\caption{The eigenvalues of the kernels of the 4-point functions in the model \eqref{H1}. The yellow curves represent the diagonal eigenvalue $k^{11}$, while the red curves are from diagonalizing the matrix of $k^{22}$, $k^{23}$, $k^{32}$ and $k^{33}$.}
				\label{fig:myev1}
				\end{figure}
				
						Diagonalizing this matrix and setting the resulting eigenvalues to 1 leads to the dimension of the operators running in the OPE channel. The eigenvalues from diagonalizing the matrix are plotted in figure~\ref{fig:myev1}. The dimension of the operators running in the diagonal $K^{11}$  is
						\bal
						h_m^0=2m+\frac{2}{3}+\g^0_m\,,\qquad m\in \mathbb{N}\,,~ \frac{1}{3}\geq\g^0_m>0\ .\label{hfr}
						\eal
						The dimensions  of the operators running in the $2\times 2$ block					fall into two towers
						\bal
						h_m^1=2m+\frac{1}{3}-\g^1_m\,,\qquad m\in \mathbb{N}\,,~ \frac{1}{3}\geq\g^1_m>0\,,\label{har}\\
						h_m^2=2m+\frac{5}{3}+\g^2_m\,,\qquad m\in \mathbb{N} \,,~ \frac{1}{3}\geq \g^2_m>0\,,\label{hsr}
						\eal
						where $\g^{0,1,2}_0=\frac{1}{3}$ and all the $\g^{0,1,2}_m$ approach to zero as $m$ increases.
						Numerically, the first few dimensions of each tower are
						\bal
						& h^0=1 ,\, 2.84 ,\,4.79 ,\,6.76,\,8.75,\,10.74 ,\ldots \\
						&  h^1= 1(0) ,\,   2.21,\, 4.30 , \,6.32 ,\, 8.32 ,\, 10.32 ,\ldots\\
						& h^2= 2 ,\,   3.82 ,\,5.78, \,7.76, \,9.74 ,\, 11.73 ,\ldots   \ .
						\eal
						Both $1$ and $0$ appears as dimensions, we choose 1 in the above list, although only the value $0$ fits the expression \eqref{har}.
						There are two dimension 1 operators that should correspond to combinations of $\f^a\f^a$ and $\bar\psi^i \psi^i$. The dimension 2 operator should correspond to the gravity mode.

Next we consider the chaotic behavior of the 4-point functions. 	To study this we need the retarded and the ``ladder rung" propagators of the vector fields in the model. Such propagators for the fermions are
derived in~\cite{Maldacena:2016hyu}, so we can simply use the results there.
But since we will need similar propagators for the bosonic field,
 we (re)derive the fermionic and the bosonic propagators,
 both as a comparison and as a double check of our computation.  	
			
In the conformal limit, correlators on the thermal circle can be obtained from that on a line by the transformation
$f(\t)=\tan(\frac{\t\p}{\b})$, so the fermionic propagator $G^\c$ and the bosonic propagators $G^\f$ at finite temperature are
  \bal
  G^\c(\t_1,\t_2)&=b^\c\text{sgn}(\t_{12})\Big(\frac{\p}{\b \sin(\frac{\p \t_{12}}{\b})}\Big)^{2\Delta_\c}\,,\quad
G^\f(\t_1,\t_2)=b^\f \Big(\frac{\p}{\b \sin(\frac{\p \t_{12}}{\b})}\Big)^{2\Delta_\f}\ .\label{gff}
\eal

The retarded propagators are defined to be the vacuum expectation value of the
\mbox{(anti-)commutators} multiplied by the step function
  \bal
 G_R^\c( t)&\equiv \< \bar\c_i( t)\c^i(0)+\c^i(0) \bar\c_i( t)\>\theta( t)\,,\\
 G_R^\f( t)&\equiv \< \f^a( t)\f^a(0)-\f^a(0) \f^a( t)\>\theta( t)\label{Gf}\ .
\eal
Comparing to the known Feynman propagators
\bal
 G^\c( t) &= \< \bar\c_i( t)\c^i(0)\>\theta( t)- \<\c^i(0) \bar\c_i( t)\>\theta(- t)\,,\\
  G^\f( t)&= \< \f^a( t)\f^a(0)\>\theta( t)+ \<\f^a(0) \f^a( t)\>\theta(- t)\,,
\eal
we obtain the retarded propagators by analytical continuations according to the known prescription (e.g.~\cite{Maldacena:2016hyu})
\begin{multline}
G_{R }^\c( t)= \frac{b^\c \p^{2\Delta_\c}}{\big(\b \sin(\frac{i \p t }{\b})\big)^{2\Delta_\c}}+\frac{b^\c \p^{2\Delta_\c}}{\big(\b \sin(-\frac{i\p  t }{\b})\big)^{2\Delta_\c}}
=\frac{2\cos({\p\Delta_\c}) b^\c   \p^{2\Delta_\c} }{\big( \b \sinh(\frac{\p  t }{\b}\big)^{2\Delta_\c}} \theta(t)\ .
\end{multline}
\begin{multline}
 G_{R }^\f( t)=   \frac{ b^\f \p^{2\Delta_\f} }{\big(\b \sin(\frac{i  \p   t }{\b}\big)^{2\Delta_\f}}- \frac{ b^\f \p^{2\Delta_\f} }{\big(\b \sin(-\frac{i  \p  t }{\b}\big)^{2\Delta_\f}}
   =   - \frac{2i \sin(\p\Delta_\f) b^\f \p^{2\Delta_\f} }{\big(  \b \sinh(\frac{\p \t }{\b}\big)^{2\Delta_\f}}\theta(t)\ .
   \end{multline}
The factor of $i$ is present because  the retarded propagator \eqref{Gf}
satisfies
\bal
G_R^{\f,\dagger}( t)=  \<[ \f^a( t)\f^a(0)]^\dagger\>\theta( t)=\< \f^a(0)\f^a( t)- \f^a( t)\f^a(0)\>\theta( t)=-G_R^\f( t)\,,
\eal
under Hermitian conjugation.
The propagators running on the ladder rung are obtained by an analytical continuation $\t\to it+\frac{\b}{2}$ from the lower half plane
\bal
G( t)_{lr}&= b \frac{\p^{2\Delta}}{\big(\b\sin(\frac{\p\t}{\b})\big)^{2\Delta}} \bigg|_{\t\to it+\frac{\b}{2}}= b \frac{\p^{2\Delta}}{\big(\b\cosh(\frac{\p t}{\b})\big)^{2\Delta}} \ .
\eal
The bosonic and the fermionic ladder rung propagators are the same since only the $\t>0$ part of the Euclidean propagator is relevant for this computation.
Furthermore, the $\l$ field remains classical in the large-$N$ limit so its ``ladder rung" propagator is again $G^\l_{lr}(t)=\frac{1}{2}$.
The retarded kernel can then be computed 						\bal
						K_R^{11}&=k^2 G^\c(t_1,t_3)G^{ \c}(t_2,t_3) (G^\l(t_3,t_4))^2 (G^\f(t_3,t_4))^2\\
						K_R^{12}&=2 k^2 G^\c(t_1,t_3)G^{ \c}(t_2,t_4) (G^\l(t_3,t_4))^2 G^\f(t_3,t_4)G^\c(t_4,t_3) \\
						K_R^{21}&=-4 k^2 G^\f(t_1,t_3)G^{\f}(t_2,t_4) (G^\l(t_3,t_4))^2 G^\f(t_3,t_4)G^\c(t_4,t_3)\\
						K_R^{22}&=-2 k^2 G^\f(t_1,t_3)G^{\f}(t_2,t_4) (G^\l(t_3,t_4))^2 (G^\c(t_4,t_3))^2\ .
						\eal
						Notice that we have included a factor $ i^2$ to all the kernels, where each $i$ factor comes from the vertex insertion in Lorentzian signature.
						
						Assuming an exponential growth ansatz
						\bal						F^\f(t_1,t_2)=\frac{e^{-\frac{\p h}{\b} (t_1+t_2) } }{( \cosh\frac{\p t_{12}}{\b} )^{\frac{1}{3}-h}}\,,\qquad F^\c(t_1,t_2)=\frac{e^{-\frac{\p h}{\b} (t_1+t_2) } }{( \cosh\frac{\p t_{12}}{\b} )^{\frac{2}{3}-h}}\,, \label{chaazt}
						\eal
the eigenvalues of the above kernels
						\bal
						&K_R^{11}F^\c(t_3,t_4)=k_R^{11}F^\c(t_1,t_2)\,,\quad K_R^{12}F^\f(t_3,t_4)=k_R^{12}F^\c(t_1,t_2)\\
						&K_R^{21}F^\c(t_3,t_4)=k_R^{21}F^\f(t_1,t_2)\,,\quad K_R^{22}F^\f(t_3,t_4)=k_R^{22}F^\f(t_1,t_2)\,,
						\eal
						can be computed using
\begin{multline}
\int_{-\infty}^{t_1} dt_3\int_{-\infty}^{t_2}  dt_4  \left(\frac{   \p }{ \b \sinh(\frac{\p  t_{13} }{\b})}\right)^{\frac{2}{q}} \left(\frac{   \p }{ \b \sinh(\frac{\p  t_{24} }{\b})}\right)^{\frac{2}{q}}\left(\frac{\p}{\b\cosh(\frac{\p t_{34}}{\b})}\right)^{2-\frac{4}{q}}\frac{e^{h \frac{\p}{\b}(t_3+t_4)/2} }{( \cosh\frac{\p t_{34}}{\b} )^{2\Delta-h}}\\
=\frac{ \Gamma \left(\frac{q-2}{q}\right)\Gamma \left(\frac{q-2}{q}\right)   \Gamma \left(\frac{2}{q}-h\right) }{\Gamma \left(-h-\frac{2}{q}+2\right)} \frac{e^{h  \frac{\p}{\b}(t_1+t_2) }}{\left(  \cosh (\frac{\p t_{12}}{\b})\right)^{\frac{2}{q}-h}}\ .\label{finitt}
\end{multline}
The results are
						\bal
						k^{11} &= \frac{ \sqrt{3}}{6 \pi   }\frac{\Gamma \left(\frac{1}{3}\right)^2 \Gamma \left(\frac{2}{3}-h\right) }{  \Gamma \left(\frac{4}{3}-h\right)}\,,\quad\qquad
						k^{12} =   \frac{ \sqrt{3}b^\c}{3 \pi b^\f}  \frac{\Gamma \left(\frac{1}{3}\right)^2 \Gamma \left(\frac{2}{3}-h\right) }{  \Gamma \left(\frac{4}{3}-h\right)}\\
						k^{21} &=    \frac{2\sqrt{3}b^\f}{3 \pi  b^\c } \frac{\Gamma \left(\frac{2}{3}\right)^2 \Gamma \left(\frac{1}{3}-h\right) }{  \Gamma \left(\frac{5}{3}-h\right)}\,,\quad
						k^{22}  =  \frac{ \sqrt{3}}{3 \pi   }\frac{\Gamma \left(\frac{2}{3}\right)^2 \Gamma \left(\frac{1}{3}-h\right) }{  \Gamma \left(\frac{5}{3}-h\right)}\ .
						\eal
The resonance of this system appears when some of the eigenvalues become 1.
						Diagonalizing this matrix, we find only one negative value
						$						h=-1$
						that makes one of the eigenvalues to be 1.
						The corresponding eigenfunctions
						$F^{\f,\c}(t_1,t_2)= \frac{e^{\frac{2\p }{\b} (t_1+t_2)/2} }{( \cosh\frac{\p t_{12}}{\b} )^{2\Delta_{\f,\c}+1}}$
						grow exponentially with the maximum Lyapunov exponent 	$ \l_L=\frac{2\p }{\b}$.

%%%%%%%%%%%%
\subsection{A slightly different model}
%%%%%%%%%%%

One could consider another model of the same set of tensor and vector fields
\bal
 H'_1 = \frac{k}{2N^2} \left( \bar{\l}^{ab}{}_{i} \bar{\l}^{eb}{}_{{j}}\f^a\f^e{\c}^i {\c}^j +{\l}^{abj} {\l}^{ebj}\f^a\f^e\bar{\c}_i \bar{\c}_j \right)\ .\label{h1p}
\eal
The corrections to the $\l$ field  are again subleading in $1/N$,
so in the $M\ll k$ limit we have $G^\l(\t_1,\t_2) =\frac{1}{2}\text{sgn}(\t_{12})$. In the low energy limit, the large-$N$ Schwinger-Dyson equations for the $\c$ and the $\f$ fields are the same as~\eqref{sd1}-\eqref{sd2}, which is solved by
\bal
G^\c(\t_1,\t_2)=\frac{b^\c\text{sgn}(\t_{12})}{|\t_{12}|^{\frac{2}{3}}}\,, \quad
G^\f(\t_1,\t_2)=\frac{b^\f }{|\t_{12}|^{\frac{1}{3}}}\,, \quad \text{ with }~ (b^\f)^2 (b^\c)^2=\frac{2\sqrt{3}}{3 \pi  k^2}\label{cfirsl}
\eal

		The 4-point function of the tensor fields is again not corrected at the leading order in $\frac{1}{N}$. The 4-point function of the vector fermions $\<\bar\c_i(\t_1)\c^i(\t_2)\bar\c_j(\t_3)\c^j(\t_4)\>$
		 mixes with the other two correlators $\<\bar\c_i(\t_1)\c^i(\t_2)\f^a(\t_3)\f^a(\t_4)\>$ and $\<\f^a(\t_1)\f^a(\t_2)\f^b(\t_3)\f^b(\t_4)\>$. The mixing is again conducted by 4 different kernels
		\bal
		K^{(11)}&=k^2 G^\c(\t_1,\t_4)G^{\c}(\t_2,\t_3) (G^\l(\t_4,\t_3))^2 (G^\f(\t_4,\t_3))^2\\
		K^{(12)}&=2 k^2 G^\c(\t_1,\t_4)G^{ \c}(\t_2,\t_3) (G^\l(\t_4,\t_3))^2 G^\f(\t_4,\t_3)G^\c(\t_3,\t_4) \\
		K^{(21)}&=4 k^2 G^\f(\t_1,\t_4)G^{\f}(\t_2,\t_3) (G^\l(\t_4,\t_3))^2 G^\f(\t_4,\t_3)G^\c(\t_3,\t_4)\\
		K^{(22)}&=2k^2 G^\f(\t_1,\t_4)G^{\f}(\t_2,\t_3) (G^\l(\t_4,\t_3))^2 (G^\c(\t_3,\t_4))^2\ .
		\eal
		There are again 3 types of eigenvectors, the same as
		the symmetric and antisymmetric fermionic $U^{s/a}$ together with a symmetric bosonic  $V$
		\bal
		U^s(\t_1,\t_2)&=\frac{1}{|\t_1-\t_2|^{\frac{2}{3}-h}}\,,\quad
		U^a(\t_1,\t_2)=\frac{\text{sgn}(\t_1-\t_2)}{|\t_1-\t_2|^{\frac{2}{3}-h}}\,,\quad
		V(\t_1,\t_2)=\frac{1}{|\t_1-\t_2|^{\frac{1}{3}-h}}\ .
		\eal
		As a result, the kernel matrix to diagonalize is again $3\times 3$, similar to that in figure~\ref{fig:ska}  but with some arrows flipped.
		The eigenvalues of the kernels
		\bal
		&K^{(11)}U^s(\t_3,\t_4)=k^{11}U^s(\t_1,\t_2)\\
		&K^{(11)}U^a(\t_1,\t_2)=k^{22}U^a(\t_1,\t_2)\,,\quad K^{(12)}V(\t_3,\t_4)=k^{23}U^a(\t_1,\t_2)\\
		&K^{(21)}U^a(\t_3,\t_4)=k^{32}V(\t_1,\t_2)\,,\quad K^{(22)}V(\t_3,\t_4)=k^{33}V(\t_1,\t_2)\,,
		\eal
		can be evaluated to be
		\bal
		k^{11}&=-\frac{2 \sqrt{3} \pi  \sin \left(\frac{1}{6} (3 \pi  h+\pi )\right) \sec \left(\frac{1}{6} \pi  (3 h+2)\right) \Gamma \left(\frac{2}{3}-h\right)}{\Gamma \left(-\frac{1}{3}\right)^2 \Gamma \left(\frac{4}{3}-h\right)}\\
		k^{22}&=-\frac{2 \sqrt{3} \pi  \cos \left(\frac{1}{6} (3 \pi  h+\pi )\right) \csc \left(\frac{1}{6} \pi  (3 h+2)\right) \Gamma \left(\frac{2}{3}-h\right)}{\Gamma \left(-\frac{1}{3}\right)^2 \Gamma \left(\frac{4}{3}-h\right)}\\
		k^{23}&=-\frac{4 \sqrt{3} \pi  \cos \left(\frac{1}{6} (3 \pi  h+\pi )\right) \csc \left(\frac{1}{6} \pi  (3 h+2)\right) \Gamma \left(\frac{2}{3}-h\right)}{\Gamma \left(-\frac{1}{3}\right)^2 \Gamma \left(\frac{4}{3}-h\right)}\\
		k^{32}&=-\frac{2 \sqrt{3} \pi  \sin \left(\frac{1}{6} \pi  (3 h+2)\right) \sec \left(\frac{1}{6} (3 \pi  h+\pi )\right) \Gamma \left(\frac{1}{3}-h\right)}{\Gamma \left(-\frac{2}{3}\right)^2 \Gamma \left(\frac{5}{3}-h\right)}\\
		k^{33}&=\frac{\sqrt{3} \pi  \sin \left(\frac{1}{6} \pi  (3 h+2)\right) \csc \left(\frac{1}{6} \pi  (8-3 h)\right) \Gamma \left(\frac{1}{3}-h\right)}{\Gamma \left(-\frac{2}{3}\right)^2 \Gamma \left(\frac{5}{3}-h\right)} \ .
		\eal

		\begin{figure}[t]
			\begin{center}
				\includegraphics[width=0.60\linewidth]{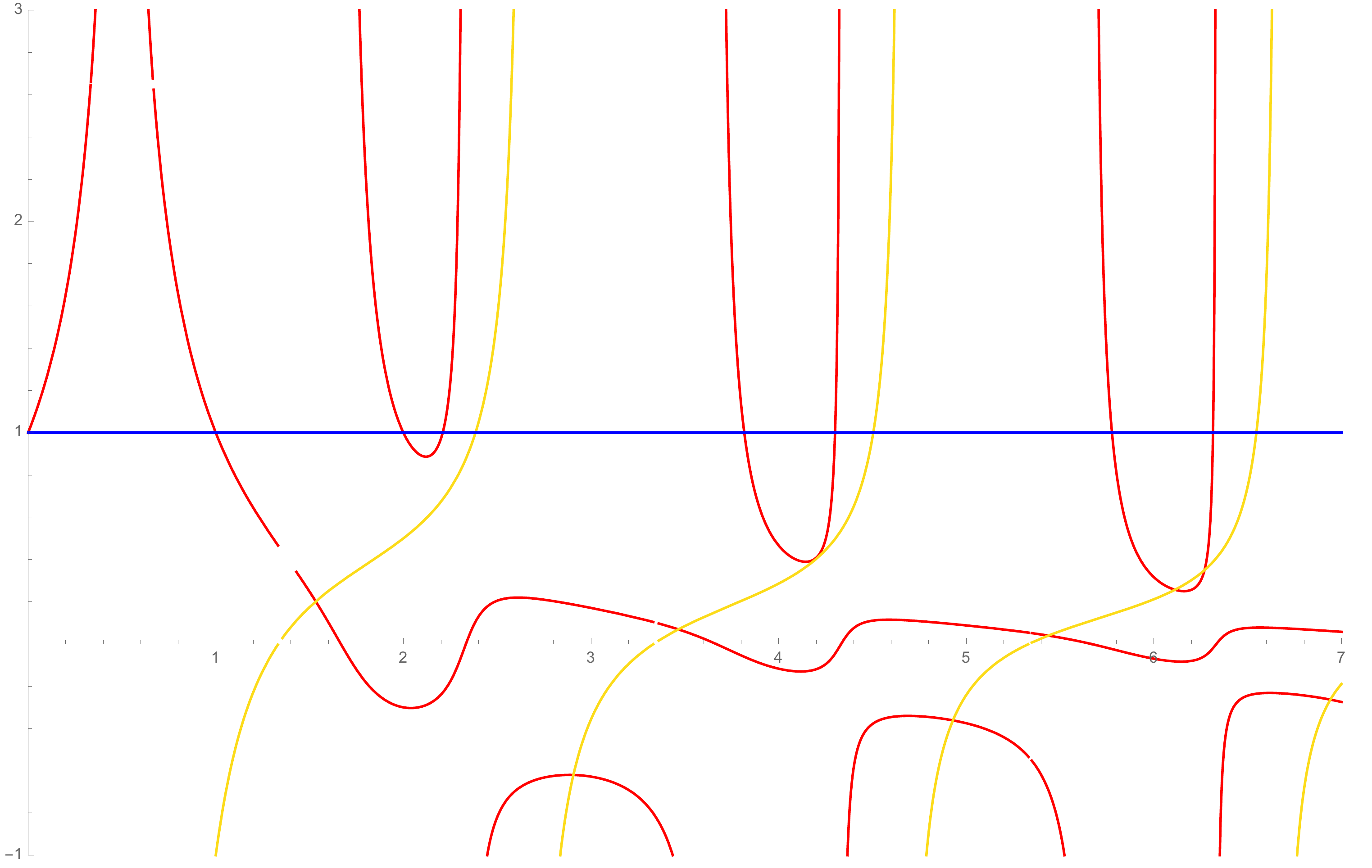}
				\end{center}
				\caption{The eigenvalues of the kernels of the 4-point functions in the model \eqref{h1p}. The yellow curves represent the diagonal $k^{11}$ eigenvalues, while the red curves are from diagonalizing the matrix of $k^{22}$, $k^{23}$, $k^{32}$ and $k^{33}$. }
				\label{fig:myev1}
				\end{figure}
				
						Diagonalizing this matrix and setting the resulting eigenvalues to 1 leads to the dimension of the operators running in the OPE channel. The eigenvalues from diagonalizing the matrix are plotted in Fig~\ref{fig:myev1}. There is a  tower of operators in the diagonal $K^{11}$ kernel with dimension
						\bal
						h_m^0=2m+\frac{2}{3}-\g'^0_m\,,\qquad m\in \mathbb{Z}_+\,, \frac{1}{3}>\g'^0_m>0\ .\label{hfr2}
						\eal
						The dimensions  of the operators running in the $2\times 2$ block fall into two towers, whose dimensions are
						\bal
						h_m^1=2m+\frac{1}{3}-\g'^1_m\,,\qquad m\in \mathbb{N}\,,~ \frac{1}{3}\geq\g'^1_m>0\,,\label{har2}\\
						h_m^2=2m+\frac{5}{3}+\g'^2_m\,,\qquad m\in \mathbb{N} \,,~ \frac{1}{3}\geq \g'^2_m>0\,,\label{hsr2}
						\eal
						where $\g'^{1,2}_0=\frac{1}{3}$ and all the $\g'^{0,1,2}_m$ approach to zero as $m$ increases.
						Numerically, the first few dimensions of these towers are
						\bal
						& h^0= 2.38 ,\,4.50 ,\,6.55,\,8.57,\,10.58  , \ldots \\
						&  h^1=  1(0),\,  2.21,\, 4.30 , \,6.32 ,\, 8.32 ,\, 10.32 ,\ldots\\
						& h^2=  2,\,  3.82 ,\,5.78, \,7.76, \,9.74 ,\, 11.73 ,\ldots   \ .
						\eal
						The operator with dimension 1 should correspond to $\f^a\f^a$. The dimension 2 operator should correspond to the gravity mode.

Next we consider the chaotic behavior of the 4-point functions. Following similar computation as in the model \eqref{H1}, we get the retarded kernels
\bal
						K_R^{11}&=k^2 G^\c(t_1,t_3)G^{ \c}(t_2,t_3) (G^\l(t_3,t_4))^2 (G^\f(t_3,t_4))^2\\
						K_R^{12}&=-2 k^2 G^\c(t_1,t_3)G^{ \c}(t_2,t_4) (G^\l(t_3,t_4))^2 G^\f(t_3,t_4)G^\c(t_4,t_3) \\
						K_R^{21}&=4 k^2 G^\f(t_1,t_3)G^{\f}(t_2,t_4) (G^\l(t_3,t_4))^2 G^\f(t_3,t_4)G^\c(t_4,t_3)\\
%						%
						K_R^{22}&=-2 k^2 G^\f(t_1,t_3)G^{\f}(t_2,t_4) (G^\l(t_3,t_4))^2 (G^\c(t_4,t_3))^2\ .
\eal
						Two more facts are taken into account to derive these retarded kernels. Firstly we have included a factor of $i^2$ to each of the kernels for the same reason as in the previous models. We have also included a factor of $-1$ to each of the kernels $K_R^{11},K_R^{21}$ for the reason explained in section \ref{mainmodel}.
						For the kernel $K_R^{12},K_R^{22}$, no factor of $-1$ is needed since the outgoing legs of the kernels are bosonic.
						
						Assuming an exponential growth ansatz
						\bal						F^\f(t_1,t_2)=\frac{e^{-\frac{\p h}{\b} (t_1+t_2) } }{( \cosh\frac{\p t_{12}}{\b} )^{\frac{1}{3}-h}}\,,\qquad F^\c(t_1,t_2)=\frac{e^{-\frac{\p h}{\b} (t_1+t_2) } }{( \cosh\frac{\p t_{12}}{\b} )^{\frac{2}{3}-h}}\,,
						\eal
the eigenvalues of the above kernels
						\bal
						&K_R^{11}F^\c(t_3,t_4)=k_R^{11}F^\c(t_1,t_2)\,,\quad K_R^{12}F^\f(t_3,t_4)=k_R^{12}F^\c(t_1,t_2)\\
						&K_R^{21}F^\c(t_3,t_4)=k_R^{21}F^\f(t_1,t_2)\,,\quad K_R^{22}F^\f(t_3,t_4)=k_R^{22}F^\f(t_1,t_2)\,,
						\eal
						can be computed by direct integrations. The results are
						\bal
						k^{11} &= \frac{ \sqrt{3}}{6 \pi   }\frac{\Gamma \left(\frac{1}{3}\right)^2 \Gamma \left(\frac{2}{3}-h\right) }{  \Gamma \left(\frac{4}{3}-h\right)}\,,\quad\qquad
						k^{12} = - \frac{ \sqrt{3}b^\c}{3 \pi b^\f}  \frac{\Gamma \left(\frac{1}{3}\right)^2 \Gamma \left(\frac{2}{3}-h\right) }{  \Gamma \left(\frac{4}{3}-h\right)}\\
						k^{21} &= -  \frac{2\sqrt{3}b^\f}{3 \pi  b^\c } \frac{\Gamma \left(\frac{2}{3}\right)^2 \Gamma \left(\frac{1}{3}-h\right) }{  \Gamma \left(\frac{5}{3}-h\right)}\,,\quad
						k^{22}  =  \frac{ \sqrt{3}}{3 \pi   }\frac{\Gamma \left(\frac{2}{3}\right)^2 \Gamma \left(\frac{1}{3}-h\right) }{  \Gamma \left(\frac{5}{3}-h\right)}\ .
						\eal
						Diagonalizing this matrix, we find only one negative value
						$						h=-1$
						that makes one of the eigenvalues to be 1.
						The corresponding eigenfunctions
						$F^{\f,\c}(t_1,t_2)= \frac{e^{\frac{2\p }{\b} (t_1+t_2)/2} }{( \cosh\frac{\p t_{12}}{\b} )^{2\Delta_{\f,\c}+1}}$
						grow exponentially with the maximum Lyapunov exponent 	$ \l_L=\frac{2\p }{\b}$.

%%%%%%%%%%%%%%%%%%%%
\section{Models with a different global symmetry}\label{uou}
%%%%%%%%%%%%%%%%%%%%

As we have discussed at the beginning of this section, it is not crucial to consider a tensor field with $O(N)\times O(N)\times U(N)$ symmetry. In the following two subsections, we consider models where the tensor field possesses a $U(N)\times O(N)\times U(N)$ symmetry.

%%%%%%%%%%%%%%%%%%%%
\subsection{A  model with a fermionic tensor field}\label{bfm}
%%%%%%%%%%%%%%%%%%%%

Since the first factor of the symmetry group becomes $U(N)$, we can consider a model similar to \eqref{h1p} but with a complex boson
\bal
 H_2 =\frac{k}{2 N^2} \Big( \bar{\l}_p{}^{b}{}_{i} \bar{\l}_q{}^{b}{}_{{j}}\f^p\f^q{\c}^i {\c}^j +{\l}^{pbi} {\l}^{qbj}\bar\f_p\bar\f_q\bar{\c}_i \bar{\c}_j \Big)\ .\label{h2p}
\eal
We will show that changing from a real boson to a comlex boson  modifies the IR physics significantly.

The Schwinger-Dyson equations for the two point functions of this model in the infrared are
\bal
 &\S^\c(\t_1,\t_2)= k^2 G^\l(\t_1,\t_2)^2\, G^{\c}(\t_1,\t_2)\, (G^{\f}(\t_1,\t_2))^2
  \\
  &\S^\f(\t_1,\t_2)= k^2 G^\l(\t_1,\t_2)^2\, (G^{\c}(\t_1,\t_2))^2\,G^{\f}(\t_1,\t_2)
  \\
 & \int d\t_2 G^\c(\t_1,\t_2)\S^\c(\t_2,\t_3)=-\delta(\t_1-\t_3)\\
 & \int d\t_2 G^\f(\t_1,\t_2)\S^\f(\t_2,\t_3)=-\delta(\t_1-\t_3)\ .
\eal
Using \eqref{freel}
and the conformal ansatz
\bal
G^\c(\t_1,\t_2)=\frac{b^\c\text{sgn}(\t_{12})}{|\t_{12}|^{2\Delta_\c}}\,, \qquad
G^\f(\t_1,\t_2)=\frac{b^\f }{|\t_{12}|^{2\Delta_\f}}\,,
\eal
the Schwinger-Dyson equations
are uniquely solved by
\bal
&\Delta_\c=\frac{1}{4}\,, \qquad \Delta_\f=\frac{1}{4}\,,\qquad (b^\f)^2 (b^\c)^2=\frac{1}{ \pi  k^2}\ .
\eal

\begin{figure}[t]
  \begin{center}
\includegraphics[width=0.60\linewidth]{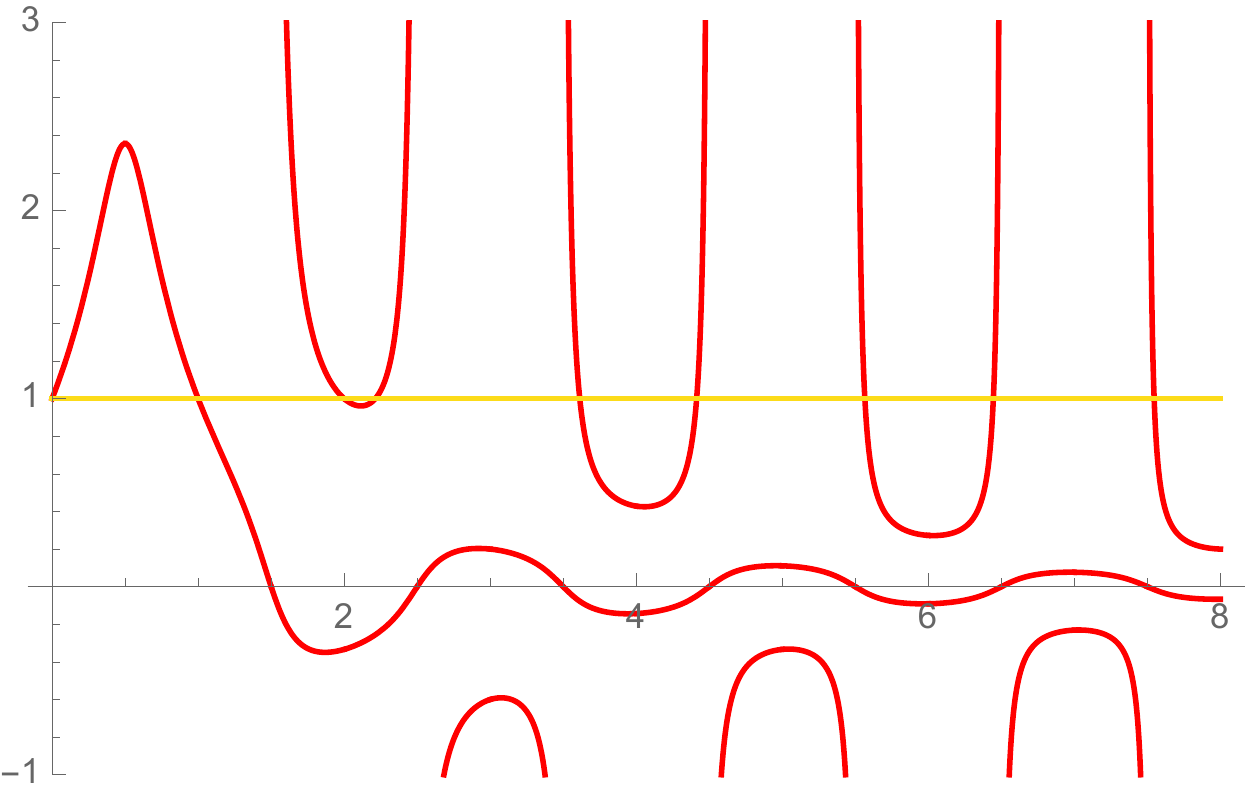}
  \end{center}
\caption{The red curve are the eigenvalues from diagonalizing the entries of the $k^{ab}$ matrix of the model \eqref{h2p}. There is another identical copy where the symmetry properties of the eigenfunctions are switched.}
\label{fig:tbf1}
\end{figure}

The 4-point functions of the vector fields are again dominated by a set of  ladder diagrams.
We consider the 4-point function
 $\<\bar\psi^i(\t_1)\psi^i(\t_2)\bar\f^p(\t_3)\f^p(\t_4)\>$, which mixes with  $\<\bar\psi^i(\t_1)\psi^i(\t_2)\bar\psi^j(\t_3)\psi^j(\t_4)\>$ and $\<\bar\f^p(\t_1)\f^p(\t_2)\bar\f^q(\t_3)\f^q(\t_4)\>$ since they share the same set of kernels.
\begin{figure}[t]
\begin{center}
\includegraphics[width=0.90\linewidth]{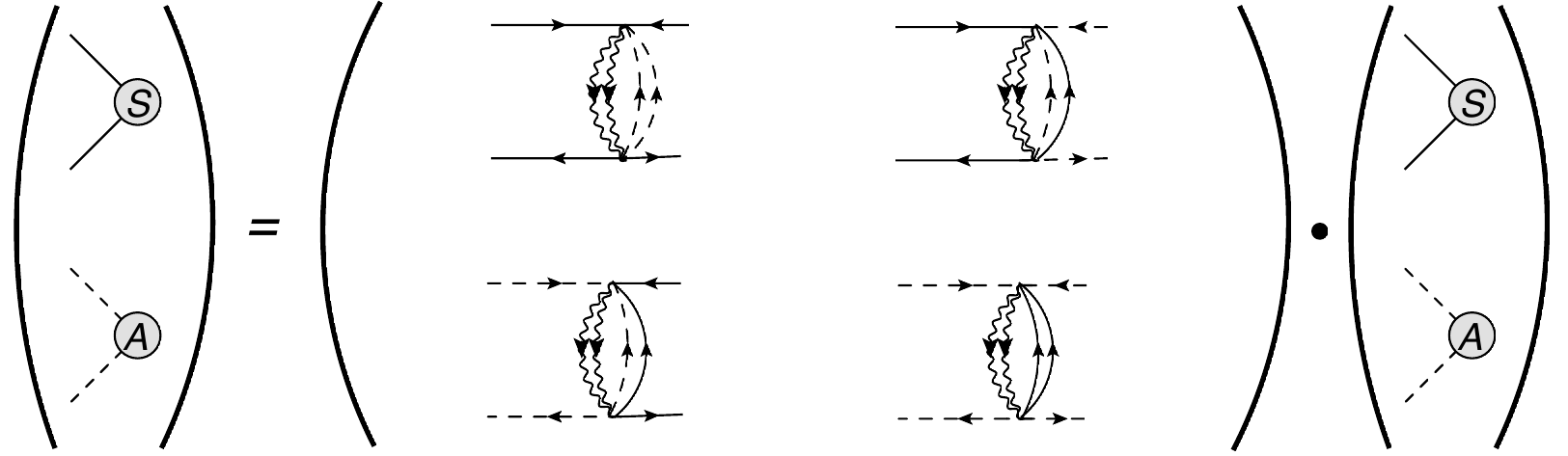}
\end{center}
\caption{Schematic form of the eigen-equations associated to the kernels of the 4-point functions in the model \eqref{h2p}.
There is another similar equation with the symmetry properties of the eigenfunctions exchanged.}
\label{fig:sk8}
\end{figure}
The 4 kernels are
\bal
K^{11}&=k^2 G^\c(\t_1,\t_4)G^{\c}(\t_2,\t_3) (G^\l(\t_4,\t_3))^2 (G^\f(\t_4,\t_3))^2\\
K^{12}&=2 k^2 G^\c(\t_1,\t_4)G^{ \c}(\t_2,\t_3) (G^\l(\t_4,\t_3))^2 G^\f(\t_4,\t_3)G^\c(\t_3,\t_4) \\
K^{21}&=2 k^2 G^\f(\t_1,\t_4)G^{\f}(\t_2,\t_3) (G^\l(\t_4,\t_3))^2 G^\f(\t_4,\t_3)G^\c(\t_3,\t_4)\\
K^{22}&=k^2 G^\f(\t_1,\t_4)G^{\f}(\t_2,\t_3) (G^\l(\t_4,\t_3))^2 (G^\c(\t_3,\t_4))^2\ .
\eal
There are $2\times 2=4$ types of eigenvectors, corresponding to symmetric/antisymmetric bosnic/fermionic ones
\bal
U^{\c/\f,s}(\t_1,\t_2)
=\frac{1}{|\t_1-\t_2|^{\frac{1}{2}-h}}\,,\qquad
U^{\c/\f,a}(\t_1,\t_2)
=\frac{\text{sgn}(\t_1-\t_2)}{|\t_1-\t_2|^{\frac{1}{2}-h}}\ .
\eal
The eigenvalues, defined by
\bal
&K^{(11)}U^{\c,s}(\t_3,\t_4)=k^{11,a}U^{\c,s}(\t_1,\t_2)\,, \quad
K^{(11)}U^{\c,a}(\t_1,\t_2)=k^{11,s}U^{\c,a}(\t_1,\t_2)\\
& K^{(12)}U^{\f,a}(\t_3,\t_4)=k^{12,a}U^{\c,s}(\t_1,\t_2)\,, \quad  K^{(12)}U^{\f,s}(\t_3,\t_4)=k^{12,s}U^{\c,a}(\t_1,\t_2)\\
&K^{(21)}U^{\c,s}(\t_3,\t_4)=k^{21,a}U^{\f,a}(\t_1,\t_2)\,, \quad K^{(21)}U^{\c,a}(\t_3,\t_4)=k^{21,s}U^{\f,s}(\t_1,\t_2)\\
&K^{(22)}U^{\f,a} (\t_3,\t_4)=k^{22,a}U^{\f,a}(\t_1,2)\,, \quad K^{(22)}U^{\f,s} (\t_3,\t_4)=k^{22,s}U^{\f,s}(\t_1,\t_2)\,,
\eal
can be evaluated explicitly in the conformal limit
\bal
&k^{11,a}=\frac{1}{2}C(h) \,, \quad
k^{12,a}= \frac{b^\c   }{ b^\f } C(h)    \,, \quad
k^{21,a}=- \frac{b^\f  }{ b^\c } S(h)  \,, \quad
k^{22,a}=-\frac{1}{2} S(h)\,,  \\
&k^{11,s}=-\frac{1}{2} S(h)\,, \quad
k^{12,s}=-\frac{b^\c  }{b^\f  }  S(h)   \,, \quad
k^{21,s}=\frac{b^\f }{ b^\c } C(h)    \,, \quad
k^{22,s}=\frac{1}{2} C(h)
\,,
\eal
where
\bal
S(h)=\frac{ 2 \cot \left(\frac{1}{4} (2 \pi  h+\pi )\right)}{   1-2h }\,, \qquad
C(h)=\frac{ 2  \tan \left(\frac{1}{4} (2 \pi  h+\pi )\right)}{ 2h-1}\ .
\eal
From these results, it is clear that the functions $\left(U^{\c,a},U^{\f,s}\right)$ close among themselves under the multiplication by the kernels. Similarly, the $\left(U^{\c,s},U^{\f,a}\right)$ closes among themselves. Consequently, we only need to diagonalize a $2\times 2$ matrix, as shown in figure~\ref{fig:sk8}, to get the eigenvalues.
Interestingly, due to the special form of the kernels, the eigenvalues corrsponding to the symmetric and antisymmetric
 eigenfunctions degenerate.
In figure~\ref{fig:tbf1}, we have plotted the eigenvalues from diagonalizing the $2\times 2$ matrix of the $\left(U^{\c,a},U^{\f,s}\right)$ eigenfunctions. The plot for the $\left(U^{\c,s},U^{\f,a}\right)$ eigenfunctions is identical.

The dimension of the operators running in the OPE channel are the values of $h$ where some eigenvalues of the kernel matrix is 1. These are the horizontal positions of the intersection between the red curves and the yellow line in figure~\ref{fig:tbf1}.
 The dimensions of the operators running in the OPE channel fall into  two sets, whose dimensions are respectively
 \bal
h_m^1&=2m+\frac{1}{2}-\h^1_m\,,\qquad m\in \mathbb{N}\,,~ \frac{1}{2}\geq \h^a_m>0\,,\label{ha}\\
h_m^2&=2m+\frac{3}{2}+\h^2_m\,,\qquad m\in \mathbb{N} \,,~ \frac{1}{2}\geq \h^2_m>0\ .\label{hs}
\eal
Both the $\h^{0,1}_m$ approach to zero as $m$ increases. A few of the lowest dimensions are
\bal
h_m^1&= 1(0),\,   2.21,\,  4.41,\, 6.45,\, 8.46,\,10.47,\,\ldots\\
h_m^2&= 2,\, 3.6,\,  5.57,\, 7.55,\,  9.54,\,11.53,\,\ldots    \ .
\eal

Notice that there is another set of antisymmetric operators that have identical dimensions as the above set of symmetric operators.
This means that there are two states for each spin with identical dimensions but different symmetry property. Naively, we guess that at each spin one of them is a bilinear of fermions and the other is a  bilinear of bosons, but it will be interesting to understand more details about this degeneracy.

Next we compute the retarded kernel to study the chaotic behavior.  Following the similar computations in the previous models, we get the retarded kernels
\bal
K_R^{(11)}&=k^2 G_R^\c(t_1,t_3)G_R^{ \c}(t_2,t_4) (G_{lr}^\l(t_3,t_4))^2 (G_{lr}^\f(t_3,t_4))^2\\
K_R^{(12)}&=-2 k^2 G_R^\c(t_1,t_3)G_R^{ \c}(t_2,t_4) (G_{lr}^\l(t_3,t_4))^2 G_{lr}^\f(t_3,t_4)G_{lr}^\c(t_3,t_4) \\
K_R^{(21)}&=2 k^2 G_R^\f(t_1,t_3)G_R^{\f}(t_2,t_4) (G_{lr}^\l(t_3,t_4))^2 G_{lr}^\f(t_3,t_4)G_{lr}^\c(t_3,t_4)\\
K_R^{(22)}&=-k^2 G_R^\f(t_1,t_3)G_R^{\f}(t_2,t_4) (G_{lr}^\l(t_3,t_4))^2 (G_{lr}^\c(t_3,t_4))^2\,,
\eal
where we have included the extra factors we discussed in the previous case.
Next we try to find eigenfunctions that grow exponentially with time.
It is easy to check that the time dependent parts in the kernels are all identical; there is only one integration to be done. Taking a similar ansatz as the \eqref{chaazt} and using again \eqref{finitt}, we find the eigenvalues of the kernels, defined by
\bal
K^{x}_R(t_1,t_2;t_3,t_4) F (t_3,t_4)&\equiv k_R^x  F (t_1,t_2)\,,
\eal
to be
\bal
k_R^{11}= \frac{ 1 }{1-2 h}\,, \quad
k_R^{12}=  \frac{-2 b^\c }{b^\f(1-2 h)}\,,\quad
k_R^{21}= \frac{ -2 b^\f }{b^\c(1-2 h)}\,,\quad
k_R^{22}= \frac{ 1 }{1-2 h}\ .
\eal
 Diagnalizing this matrix, we get two eigenvalues
 \bal
 k_{R,1}=-\frac{ 1 }{1-2 h}\,,\qquad~~ k_{R,2}=\frac{ 3 }{1-2 h}\ .
 \eal
Setting  $k_{R,1}=1$ gives
 $ h=1$,
 which leads to a decreasing eigenfunction
 with no chaotic behavior.
 On the other hand, setting $k_{R,2}=1$ renders
$ h=-1$,
 which leads to an exponentially growing eigenfunction
 ${e^{\frac{2\p }{\b} (t_1+t_2)/2} }{( \cosh\frac{\p t_{12}}{\b} )^{-\frac{3}{2} }}\,,$
 with the maximal Lyapunov exponent
$ \l_L=\frac{2\p }{\b} $.

%%%%%%%%%%%%%
\subsection{
A  model with a bosonic tensor field}\label{ffm}
%%%%%%%%%%%%%%

Given the above interesting models, we tempt to couple two different fermionic $U(N)$ models to some tensor field and check if the resulting theory is chaotic.
We thus consider a model
\bal
H_3=\frac{k}{2 N^2} \Big( \bar{\F}_p{}^{b}{}_{i} \bar{\F}_q{}^{b}{}_{{j}}\x^p\x^q{\c}^i {\c}^j +{\F}^{pbi} {\F}^{qbj}\bar\x^p\bar\x^q\bar{\c}_i \bar{\c}_j \Big)\,,\label{h3}
\eal
where $\F^{pbi}$ is a free bosonic tensor field with a propagator $G^\F(\t_1,\t_2) =1$. As in the previous sections, this propagator will not be corrected at leading order of $1/N$.
Let us emphasis again that the two vector fermions transform under two different $U(N)$ groups.
The diagrams that dominate the correlation functions of the vector fields are again melonic at the leading order of $\frac{1}{N}$. The  Schwinger-Dyson equations in the IR are
\bal
 &\S^\c(\t_1,\t_2)= k^2 G^\F(\t_1,\t_2)^2\, G^{\c}(\t_1,\t_2)\, (G^{\x}(\t_1,\t_2))^2
  \\
  &\S^\x(\t_1,\t_2)= k^2 G^\F(\t_1,\t_2)^2\, (G^{\c}(\t_1,\t_2))^2\,G^{\x}(\t_1,\t_2)
  \\
 & \int d\t_2 G^\c(\t_1,\t_2)\S^\c(\t_2,\t_3)=-\delta(\t_1-\t_3)\\
 & \int d\t_2 G^\f(\t_1,\t_2)\S^\f(\t_2,\t_3)=-\delta(\t_1-\t_3)\ .
\eal
In the conformal limit, we consider again the  ansatz
\bal
G^\c(\t_1,\t_2)&=\frac{b^\c\text{sgn}(\t_{12})}{|\t_{12}|^{2\Delta_\c}}\,,\quad G^\x(\t_1,\t_2)=\frac{b^\x \text{sgn}(\t_{12})}{|\t_{12}|^{2\Delta_\x}}\ .
\eal
The Schwinger-Dyson equations
are uniquely solved by
\bal
\Delta_\c=\frac{1}{4}\,, \qquad \Delta_\x=\frac{1}{4}\,,\qquad (b^\x)^2 (b^\c)^2=\frac{1}{4 \pi  k^2}\ .
\eal

\begin{figure}[t]
  \begin{center}
\includegraphics[width=0.60\linewidth]{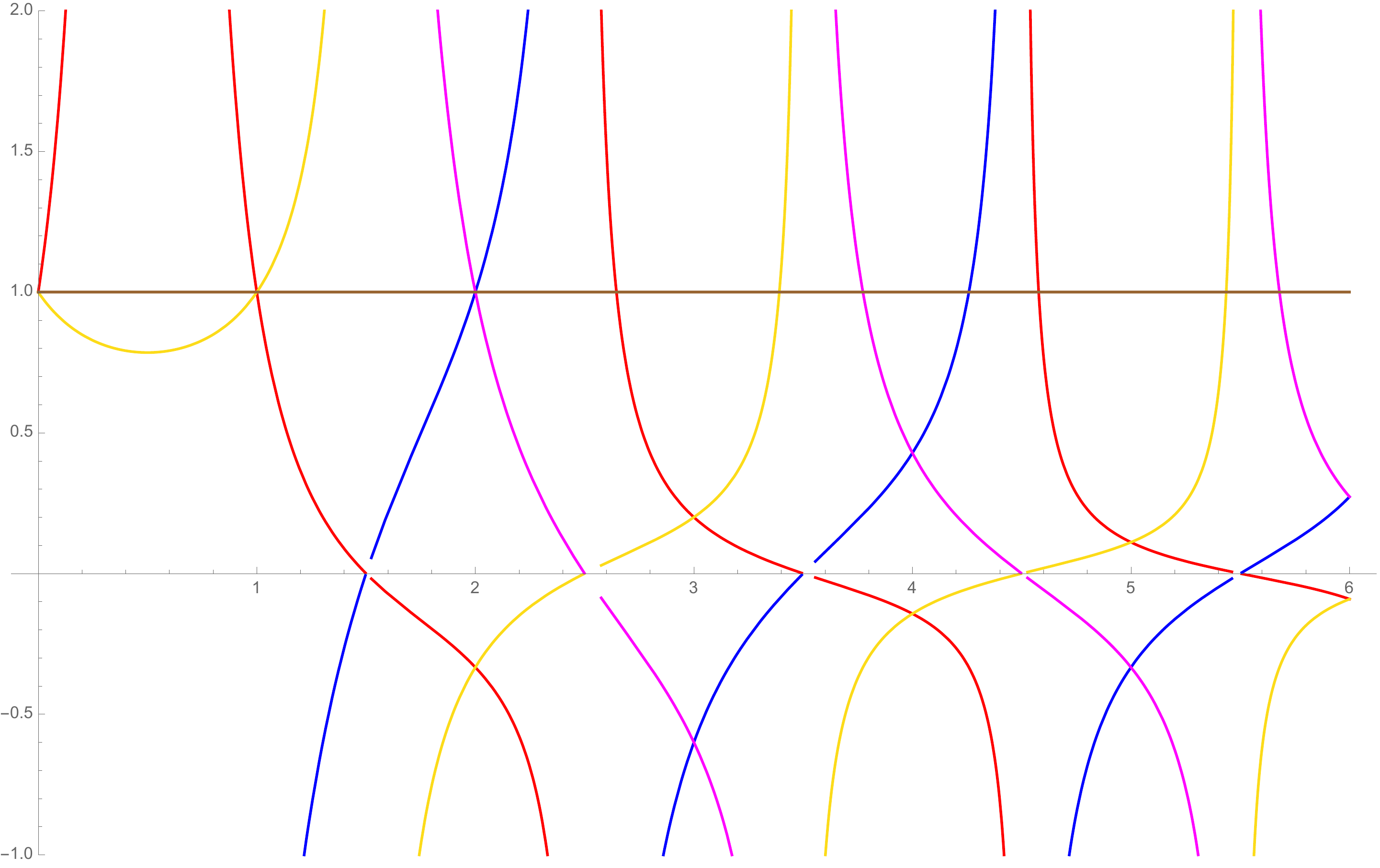}
  \end{center}
\caption{The blue, red, magenta and yellow  curves are the eigenvalues  from diagonalizing the entries of the $k^{s,1}$, $k^{s,2}$, $k^{a,1}$ and $k^{a,2}$ respectively.}
\label{fig:tff1}
\end{figure}

The 4-point function $\<\bar\x^p(\t_1)\x^p(\t_2)\bar\c^i(\t_3)\c^i(\t_4)\>$ mixes with  $\<\bar\x^p(\t_1)\x^p(\t_2)\bar\x^q(\t_3)\x^q(\t_4)\>$ and $\<\bar\c^i(\t_1)\c^i(\t_2)\bar\c^j(\t_3)\c^j(\t_4)\>$. As in the model \eqref{h2p}, there are 4 kernels involved in these correlation functions
\bal
K^{(11)}&=k^2 G^\c(\t_1,\t_4)G^{ \c}(\t_2,\t_3) (G^\l(\t_4,\t_3))^2 (G^\x(\t_4,\t_3))^2\\
K^{(12)}&=2 k^2 G^\c(\t_1,\t_4)G^{ \c}(\t_2,\t_3) (G^\l(\t_4,\t_3))^2 G^\x(\t_3,\t_4)G^\c(\t_3,\t_4) \\
K^{(21)}&=2 k^2 G^\x(\t_1,\t_4)G^{\x}(\t_2,\t_3) (G^\l(\t_4,\t_3))^2 G^\x(\t_3,\t_4)G^\c(\t_3,\t_4)\\
K^{(22)}&=k^2 G^\x(\t_1,\t_4)G^{\x}(\t_2,\t_3) (G^\l(\t_4,\t_3))^2 (G^\c(\t_3,\t_4))^2\ .
\eal
We consider 4 different eigenvectors, symmetric or antisymmetric in terms of $\x^i $, $\c^i $
\bal
U^{\c/\x,s}(\t_1,\t_2)
=\frac{1}{|\t_1-\t_2|^{\frac{1}{2}-h}}\,,\qquad
U^{\c/\x,a}(\t_1,\t_2)
=\frac{\text{sgn}(\t_1-\t_2)}{|\t_1-\t_2|^{\frac{1}{2}-h}} \ .
\eal
It is straightforward to check that the two symmetric eigenfunctions $\left(U^{\c,s},U^{\x,s}\right)$ close under the action of the kernels, and similar for the antisymmetric ones.
Furthermore, the time dependent parts  of the 4 kernels are identical, which is
\bal
K=\frac{\text{sgn}(\t_1-\t_4)}{|\t_1-\t_4|^{\frac{1}{2}}}\frac{\text{sgn}(\t_2-\t_3)}{|\t_2-\t_3|^{\frac{1}{2}}} \frac{1}{|\t_3-\t_4|}\ .
\eal
The matrix products between this kernel  and the two types of eigenfunctions are
\bal
K U^s(\t_3,\t_4)&=\frac{4\p    \tan(\p(\frac{1}{4} +\frac{h}{2} ) )}{   2h-1}    U^s(\t_1,\t_2)\\
K U^a(\t_3,\t_4)  &=  \frac{4\p    \cot(\p( \frac{1}{4}+\frac{h}{2} ))}{   2h-1}    U^a(\t_1,\t_2)\ .
\eal
Diagonalizing the (anti-)symmetric matrix of $k^{ij}$  amounts to simply diagonalizing the prefactors, whose eigenvalues are $\{\frac{3}{4\p},-\frac{1}{4\p}\}$. Consequently, the eigenvalues of the kernels of the 4-point functions in this model are
\bal
&k^{s,1}=\frac{3    \tan(\p(\frac{1}{4} +\frac{h}{2} ) )}{   2h-1}  \,,\qquad k^{s,2}=  \frac{    \tan(\p(\frac{1}{4} +\frac{h}{2} ) )}{   1-2h}  \\
&k^{a,1}=   \frac{3    \cot(\p( \frac{1}{4}+\frac{h}{2} ))}{  2h-1}  \,,\qquad k^{a,2}=    \frac{     \cot(\p( \frac{1}{4}+\frac{h}{2} ))}{  1-2h} \ .
\eal
These eigenvalues are plotted in figure~\ref{fig:tff1}.
The dimensions of the operators in the OPE channel of the above 4-point functions are
\bal
h^{s,1} &=   2,\, 4.26,\,  6.34,\, 8.38 ,\, 10.40 ,\,\ldots\,,\qquad
h^{s,2} =   1,\,  2.65,\, 4.58,\,  6.55 ,\,  8.54,\,\ldots\\
h^{a,1} &=   2,\, 3.77 ,\, 5.68,\, 7.63 ,\, 9.60 ,\,\ldots\,,\qquad ~
h^{a,2}    = 1,\, 3.39 ,\, 5.44,\, 7.45 ,\, 9.46 ,\,\ldots    \ .
\eal
Notice that there are two dimension 1, 2 operators in the OPE channels. Naively, they should be two different sets of dimension 1, 2 operators since their symmetry property are different. It is interesting to understand  which of them correspond to the gravity modes, and what are the two deformation operators.

The retarded kernels of the 4-point functions in this model are
\bal
K_R^{(11)}&=k^2 G_R^\c(t_1,t_3)G_R^{ \c}(t_2,t_4) (G_{lr}^\l(t_3,t_4))^2 (G_{lr}^\x(t_3,t_4))^2\\
K_R^{(12)}&=2 k^2 G_R^\c(t_1,t_3)G_R^{ \c}(t_2,t_4) (G_{lr}^\l(t_3,t_4))^2 G_{lr}^\x(t_3,t_4)G_{lr}^\c(t_3,t_4) \\
K_R^{(21)}&=2 k^2 G_R^\x(t_1,t_3)G_R^{\x}(t_2,t_4) (G_{lr}^\l(t_3,t_4))^2 G_{lr}^\x(t_3,t_4)G_{lr}^\c(t_3,t_4)\\
K_R^{(22)}&=k^2 G_R^\x(t_1,t_3)G_R^{\x}(t_2,t_4) (G_{lr}^\l(t_3,t_4))^2 (G_{lr}^\c(t_3,t_4))^2\ .
\eal
As we have discussed in the previous subsections, we include a factor of $-i^2$ for each of the kernels.
Assuming again an exponential  growth ansatz
\bal
F (t_1,t_2)&=\frac{e^{-\frac{\p h}{\b} (t_1+t_2) } }{( \cosh\frac{\p t_{12}}{\b} )^{2\Delta -h}}\,,
\eal
where we do  not distinguish the $\c$ or the $\x$ fields since they have the same dimension.
The eigenvalues of the above retarded kernels, defined by
\bal
K^{x}_R(t_1,t_2;t_3,t_4) F (t_3,t_4)&\equiv k_R^x  F (t_1,t_2)\,,
\eal
are
\bal
k_R^{11} =
\frac{ 1 }{  1-2h }\,,\quad
k_R^{12} =
\frac{2b^\c}{b^\x \left(1-2h\right)} \,,\quad
k_R^{21} =
\frac{2b^\x}{b^\c \left(1-2h\right)}\,,\quad
k_R^{22} =
\frac{1}{ 1-2h }\ .
\eal
Diagonalizing this matrix gives
\bal
k_{R,1}=-\frac{ 1 }{  1-2h }\,,\qquad~~ k_{R,2}= \frac{ 3 }{  1-2h }\ .
\eal
The resonance of this system appears when the eigenvalue becomes 1, which determines
 \bal
k_{R,1}=1 \quad \Rightarrow \quad h= 1\,,\qquad \qquad
k_{R,2}=1 \quad \Rightarrow \quad h=-1\ .
 \eal
Similar to the previous model, the $k_{R,1}$ eigenvalue
leads to a non-growth eigenfunction with no chaos.
 The $k_{R,2}$ eigenvalue leads to an exponential growth eigenfunction
$ {e^{\frac{2\p }{\b} (t_1+t_2)/2} }{( \cosh\frac{\p t_{12}}{\b} )^{ -\frac{3}{2}}}$, with the maximum Lyapunov exponent $\l_L=\frac{2\p }{\b} $.

    %%%%%%%%%%%%
  \section{
  A model with a self-interacting tensor field
  }\label{tself}
  %%%%%%%%%%%%%

The models discussed in the previous sections are chaotic, which originate from nontrivial dynamics of the vector fields and in particular the tensor field is classical at the leading order of $1/N$.
In this section, we try to study another chaotic model
with a self interaction of the tensor field
\bal
 H_4 =\frac{J}{2\,N^{3/2}}\,\Big( \bar{\l}^{ab}{}_{i} \bar{\l}^{ab}{}_{{j}}{\c}^i {\c}^j +{\l}^{abi} {\l}^{abj}\bar{\c}_i \bar{\c}_j \Big)+\frac{C}{2 N^{3/2}}  \bar{\l}^{ab}{}_{i} \bar{\l}^{eb}{}_{{j}}{\l}^{efi} {\l}^{afj}\ .\label{h4}
\eal
This is similar to the interaction studied in~\cite{Carrozza:2015adg,KT}, only with a slightly different symmetry group.
 The correlation functions are dominated by the set of melonic diagrams. One can sum over all melonic diagrams to get the Schwinger-Dyson equations of the 2-point function in the low energy/strong coupling limit
  \bal
  -\delta(\t_{13})&= \int d\t_2 G^\c(\t_1,\t_2)\S^\c(\t_2,\t_3)\,,\qquad  \S^\c(\t_1,t_2) =  J^2 G^\l(\t_1,\t_2)^2\, G^{\c}(\t_1,\t_2)\,,\\
  -\delta(\t_{13})&= \int d\t_2 G^\l(\t_1,\t_2)\S^\l(\t_2,\t_3)\,,\qquad   \S^\l(\t_1,\t_2) = C^2 G^\l(\t_1,\t_2)^3\ .
  \eal
Taking an ansatz
  \bal
 G^\c(\t)=\frac{b^\c \text{sgn}(\t)}{ |\t|^{1/2}}\,,\qquad G^\l(\t)=\frac{b^\l \text{sgn}(\t)}{ |\t|^{1/2}}\,,
 \eal
and making use of \eqref{todelta},  the above equations are solved by
\bal
  \Delta_\c=\frac{1}{4}\,,\qquad  \Delta_\l=\frac{1}{4}\,,\qquad
 b^\l=  \Big(\frac{1}{4\p C^2}   \Big)^{1/4}      \,, \qquad b^\c=  \Big( \frac{C^2}{4\p J^4}     \Big)^{1/4}     \ .
 \eal
The 4-point function $\<\bar\l^{ab}{}_i(\t_1){\l}^{abi}(\t_2)\bar{\l}^{de}{}_j(\t_3)\l^{dej}(\t_4)\> $ is dominated by a set of ladder diagrams with the $\l^{abi}$  fields running on the sides and the rung; the $J$ vertex does not contribute at the leading order of $1/N$.
 The kernel of the ladder diagrams is
 \bal
 K^\l(\t_1,\t_2;\t_3,\t_4)=C^2 \left(G^\l(\t_{14})G^\l(\t_{23}) G^\l(\t_{34})^2-2G^\l(\t_{13})G^\l(\t_{24}) G^\l(\t_{34})^2\right)\,,
 \eal
 which becomes
 \bal
  K^\l(\t_1,\t_2;\t_3,\t_4)=\frac{1}{4\p}\left(
  \frac{\text{sgn}(\t_{14})}{|\t_{14}|^\frac{1}{2}}\frac{\text{sgn}(\t_{23})}{|\t_{23}|^\frac{1}{2}}\frac{1}{|\t_{34}| }-2\frac{\text{sgn}(\t_{13})}{|\t_{13}|^\frac{1}{2}}\frac{\text{sgn}(\t_{24})}{|\t_{24}|^\frac{1}{2}}\frac{1}{|\t_{34}| }\right)
 \eal
 in the IR.
There are two sets of eigenfunctions since the model is complex
\bal
U_h^s(\t_1,\t_2)& =\frac{1}{|\t_1-\t_2|^{\frac{1}{2}-h}}\,\qquad
U_h^a(\t_1,\t_2) =\frac{\text{sgn}(\t_1-\t_2)}{|\t_1-\t_2|^{\frac{1}{2}-h}}\ .\label{azt0a}
\eal
The associated eigenvalues can be evaluated directly
\bal
&K^\l(\t_1,\t_2;\t_3,\t_4)U_h^s(\t_3,\t_4)=k_h^{\l,s} U_h^s(\t_1,\t_2)\\
&K^\l(\t_1,\t_2;\t_3,\t_4)U_h^a(\t_3,\t_4)=k_h^{\l,a} U_h^a(\t_1,\t_2)  \ .
\eal
where
\bal
k_h^{\l,s} &=-\frac{     \cot(\p(\frac{1}{4}-\frac{h}{2}))}{2h-1} \,,\qquad~~
k_h^{\l,a} =  \frac{3    \tan(\p(\frac{1}{4}-\frac{h}{2}))}{ 2h-1}\ . \label{tensoregv}
  \eal
 The dimensions of the set of operators running in the OPE channel are obtained by setting these eigenvalues to one.
  Numerically, the dimensions are
  \bal
  k^{\l,a}=1&\quad \Rightarrow \quad h^{\l,a}=2,\, 3.77 ,\, 5.68,\, 7.63 ,\, 9.60 ,\,\ldots \label{kla}\\
   k^{\l,s}=1&\quad \Rightarrow \quad h^{\l,s}=1,\, 2.65,\, 4.58,\, 6.55 ,\, 8.54,\,\ldots\ .\label{kls}
  \eal
These dimensions
are the same as the uncolored tensor model with $U(N)\times O(N)\times U(N)$ symmetry discussed in~\cite{KT}.

\begin{figure}[t]
 \begin{center}
\includegraphics[width=0.60\linewidth]{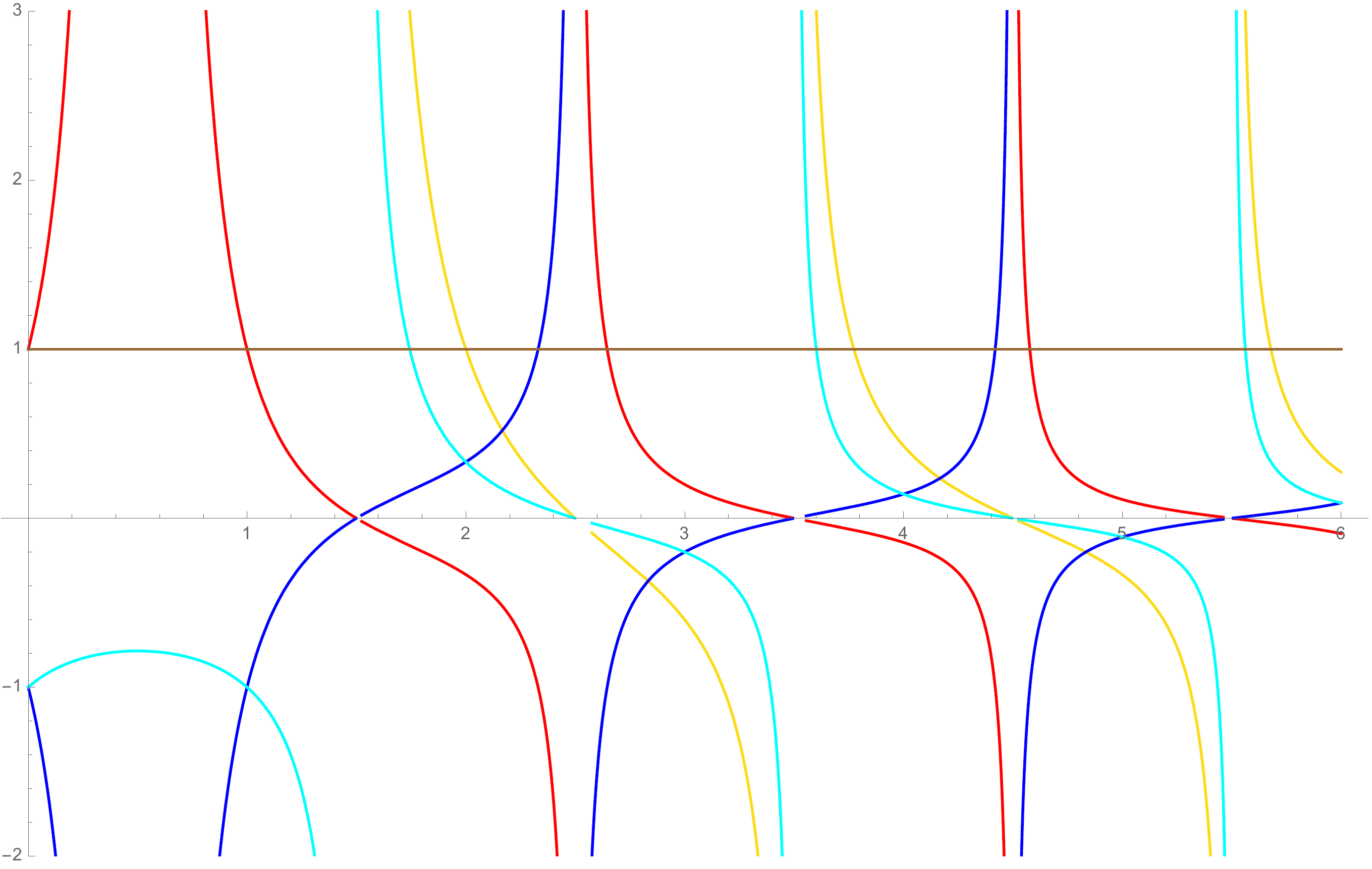}
 \end{center}
\caption{The eigenvalues of the kernels of the 4-point functions in the model \eqref{h4}. The red and yellow curves are the eigenvalues of the symmetric and antisymmetric eigenfunctions of the kernel of the tensor 4 point function. The blue and cyan curves correspond to the symmetric and antisymmetric eigenvalues of the vector kernel.}
\label{fig:vec1}
\end{figure}

\begin{figure}[t]
  \begin{center}
\includegraphics[width=0.55\linewidth]{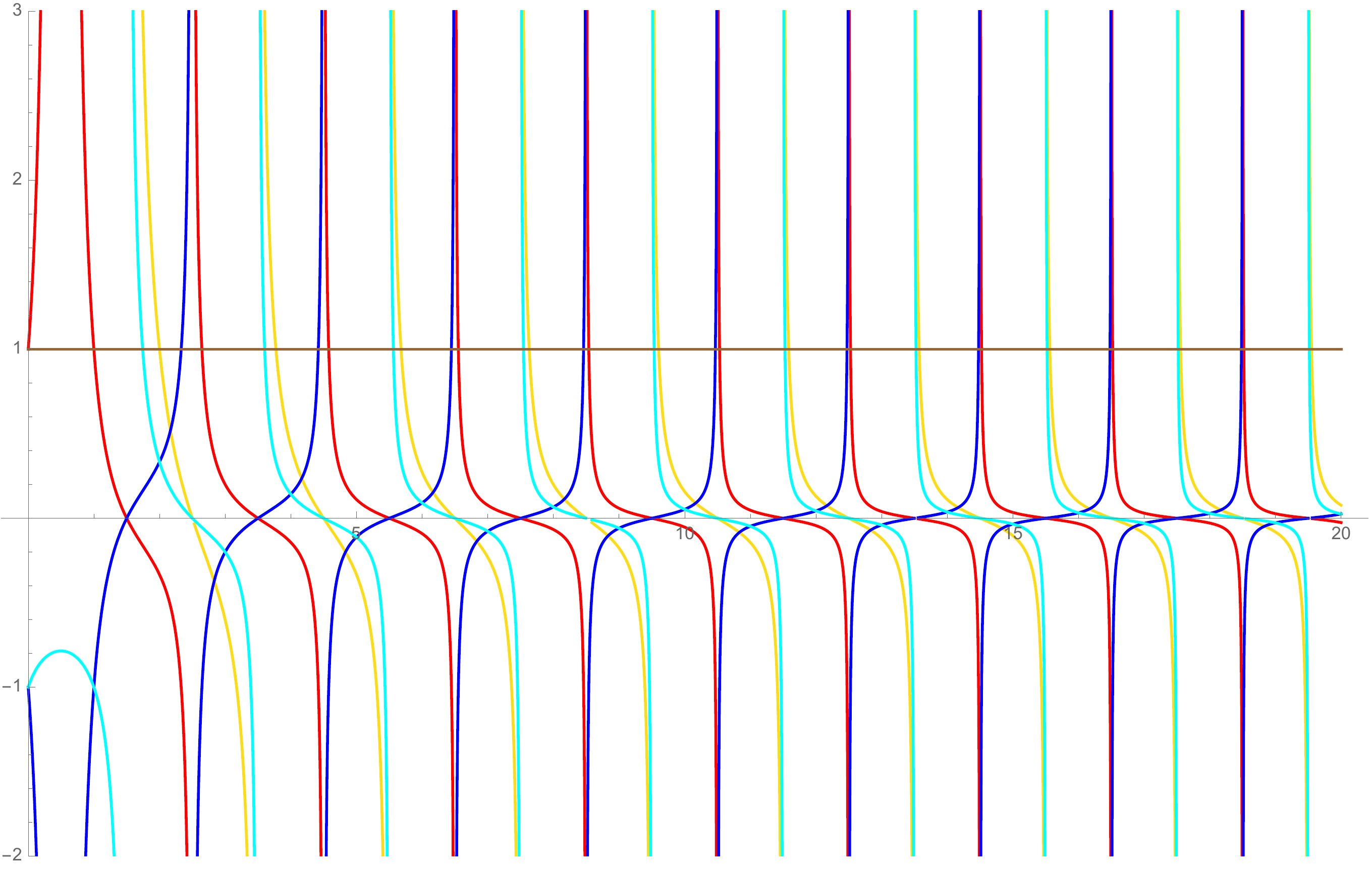}
  \end{center}
\caption{The dimensions of the operator in the vector and tensor 4-point function asymptote to the same value as the dimension increases.}
\label{fig:vec2}
\end{figure}

The 4-point function of the vector fields $\<\bar\c^i(\t_1)\c_i(\t_2)\bar\c^j(\t_3)\c_j(\t_4)\>$ can be computed by summing over ladder diagrams again of the type shown in figure~\ref{fig:ladder1}.
The kernel of these ladder diagrams is
\bal
K^\c(\t_1,\t_2;\t_3,\t_4)&=   J^2 G^\c(\t_{14})G^\c(\t_{23}) \left(G^\l(\t_{34}) \right)^2\ .
\eal
This kernel again admits symmetric and antisymmetric eigenfunctions whose numerical forms are identical to \eqref{azt0a}. The corresponding eigenvalues can be computed similarly
\bal
&K^\c(\t_1,\t_2;\t_3,\t_4)U_h^s(\t_3,\t_4)=k^{\c,s} U_h^s(\t_1,\t_2)\\
&K^\c(\t_1,\t_2;\t_3,\t_4)U_h^a(\t_3,\t_4)=k^{\c,a} U_h^a(\t_1,\t_2)  \,,
\eal
where
\bal
k^{\c,s}    =   \frac{     \cot(\p(\frac{1}{4}-\frac{h}{2}))}{ 2h-1}  \,,\qquad ~~
k^{\c,a} =  \frac{    \tan(\p(\frac{1}{4}-\frac{h}{2}))}{  2h-1} \ .
\eal
They are related to \eqref{tensoregv} by
\bal
k_h^{\c,s} =-k_h^{\l,s}\,,  \qquad k_h^{\c,a} =\frac{k_h^{\l,a} }{3}\ .\label{vectoregv}
\eal
The eigenvalues \eqref{tensoregv} and \eqref{vectoregv} are plotted in figure~\ref{fig:vec1}.
The dimensions of the set of operators running in the OPE channel are obtained from setting the eigenvalues \eqref{vectoregv} to one. Numerically, the dimensions are
  \bal
&  k^{\c,a}=1\quad \Rightarrow \quad h^{\c,a}=1.74\,, 3.60\,, 5.56\,, 7.55\,, 9.54\,,\ldots\label{kca}\\
&   k^{\c,s}=1\quad \Rightarrow  \quad h^{\c,s}=2.33\,, 4.42\,, 6.45\,, 8.46\,, 10.47\,,\ldots\ .\label{kcs}
  \eal
These are two different towers of operators comparing to \eqref{kla} and \eqref{kls}. The dimensions \eqref{kca} and \eqref{kcs} approach to the dimensions \eqref{kla} and \eqref{kls} respectively as the dimension increases. This is shown in figure~\ref{fig:vec2}.
Moreover, the operators with dimension 1 and 2 do not run in the OPE channel of the 4-point function of the vector fields. This could mean that the AdS$_2$ graviton only contributes to the correlation functions of the vector fields  in the bulk at loop level, since the above result is at the leading order of $1/N$. This gives a hint of the form of the coupling between gravity and the dual of the vector fields: the gravity fields appears at least quadratically in the vertex. It will be interesting to have a better understanding of this result.

Next we consider the out-of-time-order 4-point functions and check if the model \eqref{h4} is chaotic. For this we study retarded kernels and check if it has any eigenfunction that grows exponentially in time.
We find the two retarded kernels to be
\bal
 K_R^\l( t_1, t_2; t_3, t_4)
 &= 3C^2 G_R^\l(t_{13})G_R^\l(t_{24}) G_{lr}^\l(t_{34})^2\label{rkl}\\
  K_R^\c(t_1,t_2;t_3,t_4)&=    J^2 G_R^\c(t_{13})G_R^\c(t_{24}) \left(G_{lr}^\l(t_{34}) \right)^2 \,,\label{rkc}
 \eal
 where we have included appropriate factors for different terms as we has discussed in previous sections.
An exponential  growth ansatz
\bal
F^x(t_1,t_2)&=\frac{e^{\l_L (t_1+t_2)/2} }{( \cosh\frac{\p t_{12}}{\b} )^{2\Delta_x-h}}\,,\qquad x=\{\l,\c\}\label{gazt}
\eal
becomes an eigenfunction of the retarded kernel \eqref{rkl}
\bal
& K_R^\l( t_1, t_2; t_3, t_4)F^\l(t_1,t_2)=k_R^\l F^\l(t_1,t_2)\,,
\eal
when
\bal
&k_R^\l=     \frac{3}{  (1-2h) }\,,\qquad \text{ and } \quad \l_L=-\frac{2\p h}{\b}\ .
\eal
This can be verified explicitly using \eqref{finitt}.
The resonance of this system appears when the eigenvalue becomes 1, which happens at
$ h=-1$.
This leads to an exponentially growing eigenfunction
with the maximal Lyapunov exponent
$\l_L=\frac{2\p }{\b}$.
On the other hand, the eigenvalue  of the  4-point function~$\<\bar\c^i(t_1)\c_i(t_2)\bar\c^j(t_3)\c_j(t_4)\>$ is
\bal
k_R^\c=     \frac{ 1 }{1-2h} \,,\qquad \text{ and } \quad \l_L=-\frac{2\p h}{\b}\ .
\eal
The eigenvalue becomes 1 at
$ h=0$. This leads to a non-growing eigenfunction
$ F^\c(t_1,t_2)=( \cosh\frac{\p t_{12}}{\b} )^{ -\frac{1}{2}} $ without quantum chaos.

  As a result, the chaotic behavior only appears in the 4-point function of the tensor fields. This is consistent with the fact that the $h=2$ mode, which corresponds to the gravity mode, only appears in the OPE channel of the 4-point function of the tensor fields.
  This suggests that the holographic dual of the tensor field in this model contains the dilaton-gravity sector on AdS$_2$, which is propably identical to the dual of the SYK$_4$ model. 

One more comment is that the nature of this perturbation \eqref{h4} is different from the perturbations discussed in the previous sections. This perturbation gives nontrivial correction to the tensor field that has $N^3$ degrees of freedom, which will clearly dominate the vector fields in the large-$N$ limit. Therefore the deformation \eqref{h4} leads to a chaotic behavior by changing the dominant field. This is very different from the deformations in the previous sections, where the dominant degrees of freedom are from the vector fields both before and after the perturbation is turned on.

%%%%%%%%%%%%%%%
\section{Conclusion}
%%%%%%%%%%%

In this paper we try to explore relations between SYK type solvable models and weakly coupled vector models. The relation is demonstrated by a simple toy model \eqref{action} that couple a tensor field with a vector field.
If the tensor field is heavy,  we can safely integrate it out,
and the resulting theory is the 1-dimensional Gross-Neveu model.
If instead the scale of the interaction is much larger than the mass of the tensor,
at low energy the theory behaves like an SYK$_2$ model. We can turn on a marginally relevant deformation at this energy scale and the theory flows to a new SYK-like fix point that is chaotic. If the sign of the perturbation is flipped, it is marginally irrelevant and the theory returns to the unperturbed model. Therefore a phase transition is associated to the sign flip of the perturbation.
Other deformations of the model \eqref{action} exist with well defined large-$N$ limits.
We study some of these deformations and the resulting theories are all chaotic. None of the models we considered have disorder, and the global symmetries are explicit. So one can gauge the $O(N)$ or $U(N)$ symmetries and look for bulk dual of only the gauge invariant operators.

The SYK model contains a tower of operators in the low energy limit. They are believed to be dual to the tower of states in some 2D string theory where there is only field at each spin due to the absence of the perpendicular degrees of freedom.
On the other hand, some of the models we discussed, e.g. \eqref{H1}, \eqref{h1p}  and \eqref{h3}, contain more than one tower of operators.
Thus a natural question is which, if any, tower corresponds to the 2D string modes, and what is the interpretation of the other operators.

\acknowledgments

We are indebted to Wenbo Fu,  Frank Ferrari and Grigory Tarnopolsky for many illuminating conversations, and especially to Marcus Spradlin and Anastasia Volovich for collaborations during the early stage of this project.
We thank Nikolay Bobev, Mark Henneaux, Enrico Herrmann, Antal Jevicki, Alexei Kitaev, Ritam Sinha, Douglas Stanford, Kenta Suzuki, Jacobus Verbaarschot and Ida Zadeh
for helpful discussions on related topics, and especially grateful to  Marcus Spradlin and Douglas Stanford for important comments on
the draft.
We appreciate the warm hospitality by ETH Zurich, KU Leuven, the Universit\'{e} Libre de Bruxelles and Galileo Galilei Institute for Theoretical Physics during various stages of this work.
This work was supported by the US Department of Energy under contract DE-SC0010010 Task A and partially by a  Young Investigator Training Program fellowship from INFN during the completion of this work.

\end{document}